\documentclass[aps,prx,twocolumn,showpacs,amsmath,amssymb,floatfix,superscriptaddress]{revtex4-1}

\usepackage[outline]{contour}

\usepackage{color}
\usepackage{bbm}
\usepackage{graphicx}

\usepackage{dcolumn}
\usepackage[utf8]{inputenc}
\usepackage{graphicx}
\usepackage{epstopdf}
\usepackage{amsthm}

\usepackage{amsmath}
\usepackage{empheq}

\usepackage{eufrak}
\usepackage{bbm}
\usepackage{braket}
\usepackage{amssymb}
\usepackage{mathtools}
\usepackage[usenames,dvipsnames]{xcolor}
\usepackage{tikz}
\usepgflibrary{shapes.arrows}
\usetikzlibrary{calc, positioning, shapes.arrows}

\usepackage{cases}
\usepackage{smartdiagram}
\usepackage{latexsym}
\usepackage[colorlinks=true,citecolor=Cerulean,linkcolor=RubineRed,urlcolor=Cerulean]{hyperref}

\usepackage{cleveref}

\contourlength{0.5pt}
\contournumber{20}

\definecolor{S_Blue}{RGB}{0,135,252}
\definecolor{S_Red}{RGB}{214,13,63}
\definecolor{Blue}{RGB}{47,89,151}
\definecolor{S_Grey}{RGB}{150,150,158}
\definecolor{S_Yel}{RGB}{255,204,0}

\definecolor{S_Green}{RGB}{102,204,0}
\definecolor{S_Brown}{RGB}{154,41,41}

\DeclareFontFamily{OT1}{pzc}{}
\DeclareFontShape{OT1}{pzc}{m}{it}{<-> s * [1.15] pzcmi7t}{}
\DeclareMathAlphabet{\mathpzc}{OT1}{pzc}{m}{it}

\graphicspath{ {figures/figure1/}{figures/figure2/} }

\DeclareMathOperator{\sign}{sign}

\newcommand{\qs}{ \textnormal{speedup} }
\newcommand{\dota}{ \mathpzc{\dot a}}

\newcommand{\fancya}{\mathpzc{a}}

\newcommand{\info}{\mathcal{I}_F} 

\newcommand{\slog}{ {L}} 
\newcommand{\slogq}{ {L}_{C}} 
\newcommand{\slogc}{ {L}_{I}} 

\newcommand{\dt}{\frac{d}{dt}} 

\newcommand{\id}{\mathbbm{1}} 
\newcommand{\tr}[1]{\operatorname{\textnormal{Tr}}\left( {#1} \right)} 
\newcommand{\trs}[2]{\operatorname{\textnormal{Tr}}_{{#1}}\left( {#2} \right)}  
\newcommand{\cov}{\textnormal{cov}}

\newcommand{\sys}{\mathcal{S}} 
\newcommand{\env}{\mathcal{E}} 

\begin{document}


\title{
Unifying Quantum and Classical Speed Limits on Observables 
 }

\author{Luis Pedro Garc\'ia-Pintos}
\thanks{Corresponding author}
\email{lpgp@umd.edu}
\affiliation{Joint Center for Quantum Information and Computer Science and Joint Quantum Institute, NIST/University of Maryland, College Park, Maryland 20742, USA}

\author{Schuyler Nicholson}
\affiliation{Department of Chemistry, Northwestern University, 2145 Sheridan Road, Evanston, Illinois 60208, USA}

\author{Jason~R.~Green}
\affiliation{Department of Chemistry,\
  University of Massachusetts Boston,\
  Boston, MA 02125, USA
}
\affiliation{Center for Quantum and Nonequilibrium Systems,\
  University of Massachusetts Boston,\
  Boston, MA 02125, USA
}
\affiliation{Department of Physics,\
  University of Massachusetts Boston,\
  Boston, MA 02125, USA
}

\author{Adolfo del Campo}
\affiliation{Department  of  Physics  and  Materials  Science,  University  of  Luxembourg,  L-1511  Luxembourg, G. D.  Luxembourg}
\affiliation{Donostia International Physics Center,  E-20018 San Sebasti\'an, Spain}
\affiliation{Department of Physics,\
  University of Massachusetts Boston,\
  Boston, MA 02125, USA
}

\author{Alexey V. Gorshkov}
\affiliation{Joint Center for Quantum Information and Computer Science and Joint Quantum Institute, NIST/University of Maryland, College Park, Maryland 20742, USA}


\begin{abstract}
The presence of noise or the interaction  with an environment can radically change the dynamics of observables of an otherwise isolated quantum system. We derive a bound on the speed with which observables of open quantum systems evolve. This speed limit divides into Mandalestam and Tamm's original time-energy uncertainty relation and a time-information uncertainty relation recently derived for classical systems, generalizing both to open quantum systems. By isolating the coherent and incoherent contributions to the system dynamics, we derive both lower and upper bounds to the speed of evolution. We prove that the latter provide tighter limits on the speed of observables than previously known quantum speed limits, and that a preferred basis of \emph{speed operators} serves to completely characterize the observables that saturate the speed limits. We use this construction to bound the effect of incoherent dynamics on the evolution of an observable and to find the Hamiltonian that gives the maximum coherent speedup to the evolution of an observable.  
\end{abstract}

\maketitle

\section{Introduction}
\label{sec:introduction}
Quantum speed limits are bounds on the rate of evolution of quantum systems~\cite{mandelstamtamm1945}. 
Mandelstam and Tamm first derived 
their bound on speed
for systems evolving unitarily under a Hamiltonian $H$~\cite{mandelstamtamm1945}. 
They proved that the rate of change of the expectation value $\langle A \rangle$
of an arbitrary observable $A$ satisfies
\begin{align}
\label{eq:MT}
\left| \frac{d \langle A \rangle}{dt} \right| \leq 2 \Delta A \Delta H,
\end{align}
where $\Delta A \coloneqq \sqrt{ \langle A^2 \rangle - \langle A \rangle^2}$ and $\Delta H $ are the standard deviations of the observable and the Hamiltonian, respectively (units are such that $\hbar = 1$). 
This general result bounds the speed of evolution of any physical quantity of an isolated quantum system. 
Mandelstam and Tamm further considered the projection onto the initial state of the system, $A \coloneqq \ket{\psi_0}\!\bra{\psi_0}$, as an observable of interest. Equation~\eqref{eq:MT} then implies a bound on how fast the \emph{state} of the system evolves. 
They proved that the minimum time $\tau^\perp$ for a system to evolve between two orthogonal states satisfies $\tau^\perp  \geq \pi/(2\Delta H)$. 
This gives an ultimate limit to the speed of evolution in the system: a minimum time has to elapse  for the state of the system to evolve into a distinguishable state, $t \geq \tau^\perp$ .

Since then, the focus on state distinguishability instead of observables has been predominant,  with most works adopting a metric in Hilbert space and deriving bounds on its rate of change.
For instance, some have focused on
alternative bounds to that of Mandelstam and Tamm~\cite{margoluslevitin1998,Zielinski06,
FrowisPRA2012,toth2014quantum}, involving various metrics~\cite{Uhlmann92,PATI199540,
MacconePRA2003,Zhang14,PiresPRX2016,
MarvianPRA2016,Deffner_2017,Campaioli18,
Campaioli2019tightrobust,Sun19,sun2015quantum,YuanPRR2020} and covering more general dynamical regimes~\cite{DavidovichPRL2013,
delCampoPRL2013,DeffnerLutzPRL2013,LPGPcontQSL2019}.
In this way, quantum speed limits have found applications to
a range of topics, 
 including   quantum control~\cite{CanevaPRL2009,Campbell17,Funo17}, limits to computation~\cite{margoluslevitin1998,LloydNature2000,
 LloydPRL2002}, parameter estimation \cite{braunstein1996generalized,Giovannetti11,Beau17}, quantum thermodynamics~\cite{delcampo14,Campaioli17}, quantum annealing~\cite{Suzuki20}, quantum information theory~\cite{jing2016fundamental,
 Deffner20,Pires20,Campaioli20}, 
  as well as the dynamics of many-body~\cite{CampbellPRL2020,delcampo20} and open quantum systems~\cite{Marvian15}. They have even been extended to classical settings by using metrics in the space of probability distributions~\cite{margolus2011finitestate,Shanahan18,
 Okuyama18,HasegawaPRE2020}. 
Some studies have deviated from the focus on speed limits in Hilbert space, deriving bounds 
on quantum thermodynamic processes~\cite{ito2017fundamental,juliafarre2018bounds,
garcapintos2019fluctuations}.
However, metric-based speed limits remain prevalent~\cite{DeffnerReview17}.

Approaches based on metrics in Hilbert space have a shortcoming: while any pair of orthogonal states are distinguishable under \emph{some} measurement, oftentimes relevant observables remain unchanged. For example, consider an ensemble of two-level systems that evolve from state $\ket{\uparrow \uparrow \uparrow \dots \uparrow \uparrow}$ to a state $\ket{\uparrow \uparrow \uparrow \dots \uparrow \downarrow}$ over a time $\tau^\perp$.
 All distances between these states achieve their maximum value, but interesting observables such as the magnetizations $\sum_j\sigma_j^z$ or $\sum_j \sigma_j^x$ barely change (here $\sigma^\alpha$ are Pauli matrices).
 Observables can be thought of as `filter functions', sensitive only to restricted parts of the dynamics of the state.
While bounds on $\tau^\perp$ provide information about the fastest evolving Hermitian operators, they may not reflect the dynamics of experimentally relevant physical observables~\footnote{An interesting example where speed limits in Hilbert space correctly capture the dynamics of a many-body system is studied in~\cite{CampbellPRL2020}}.
This situation is exacerbated by the fact that the speed limits of different metrics can vary significantly~\cite{PiresPRX2016,Campaioli18}. 
We see this as a prime example that  highlights the need for speed limits on physically-grounded figures of merit, and perhaps, of direct relevance to experimental measurements.

In this work, we address the shortcomings of derivations of speed limits for distances in Hilbert space. We do so by deriving limits on the evolution of arbitrary observables for arbitrary differentiable dynamics. 
This extends the original derivation by Mandelstam and Tamm, which was restricted to isolated systems evolving unitarily. 
In doing so, we also generalize Mandelstam and Tamm's version of the quantum time-energy uncertainty relation to open quantum systems. 
This result is the quantum analog of a \emph{time-information uncertainty relation} recently introduced for stochastic dynamics of classical systems~\cite{nicholson2020timeinformation},
in which the maximal rate of change of an observable depends on (classical) fluctuations in the observable and fluctuations in the surprisal rate of the system.

The remainder of the paper is organized as follows. Section~\ref{sec:generalQSL} shows a general bound on the rate of change of observables for open quantum systems in terms of the quantum Fisher information and discusses connections to the quantum Cram\'er-Rao bound. 
The core results of this paper are contained in
Section~\ref{sec:coherent-incoherent}, where we derive strictly tighter upper bounds than those in Sec.~\ref{sec:generalQSL} by decomposing the dynamics of a quantum system in terms of coherent and incoherent contributions. This technique also allows us to derive, for the first time, lower bounds on the speed of observables.
We show in Sec.~\ref{sec:boundsfisher} how to connect these bounds to energy uncertainties of the system and derive bounds on energy and entropy rates, providing a simple proof of the `small incremental entangling theorem'. 
Section~\ref{sec:saturation} focuses on the tightness of the speed limits obtained in Secs.~\ref{sec:generalQSL} and~\ref{sec:coherent-incoherent}. 
 We derive limits to the total change of an observable in Sec.~\ref{sec:geometry}, where we use them to bound the integrated effects from incoherent dynamics on an open quantum system. In Sec.~\ref{sec:discussion}, we provide conclusions and a discussion. Finally, in the Appendixes, we present details omitted in the main text.

\section{Speed limits on observables in open quantum systems}
\label{sec:generalQSL}
Let $\rho_t$ denote an arbitrary density matrix of a quantum system, possibly characterizing a statistical mixture over pure states.
Its dynamics can be formally expressed by the \emph{symmetric logarithmic derivative} $\slog$, implicitly defined by $ \tfrac{d \rho_t}{dt} \eqqcolon \tfrac{1}{2}\{ \slog,\rho_t \} $~\cite{helstrom1969quantum,
holevo2011probabilistic,
BraunsteinCaves1994,
 braunstein1996generalized}, where $\{A,B \} \nobreak =  \nobreak AB+BA$ denotes an anticommutator. 
Using this equation of motion, we show (Appendix~\ref{app:qPrice}) that  
the expectation value $\langle A \rangle = \tr{A\rho_t}$  
evolves according to a generalized form of the Ehrenfest equation, 
\begin{align}
\label{eq:generaldyn}
\frac{d\langle A \rangle}{dt} &= \cov\left( \slog,A \right) + \left\langle \frac{dA}{dt} \right\rangle \eqqcolon  \dota + \left\langle \frac{dA}{dt} \right\rangle,
\end{align}
where $\cov(A,\slog) \nobreak \coloneqq \nobreak \frac{1}{2}  \tr{\rho_t \{A,\slog\}} - \langle A \rangle \langle \slog \rangle$ is the symmetrized covariance.
Here, we distinguish between the term 
$ \left\langle \tfrac{dA}{dt} \right\rangle $ that stems from any explicit time dependence of the observable and
$\dota \nobreak \coloneqq \nobreak\tr{A \tfrac{d\rho_t}{dt} } \nobreak = \nobreak \, \cov\left( \slog,A \right)$, which depends on the rate of change of the state.
Note that for time-dependent observables $\fancya$ is not a state function but is instead 
defined through 
the path-dependent integral $\fancya \coloneqq \int\dota \, dt$. An analogy is that of heat and work in thermodynamics, whose total changes are defined via the infinitesimal changes on a given process and cannot generally be defined only in terms of the initial and final states. In fact, if $A = H(t)$ is the Hamiltonian of the system, $\dota$ and $ \langle \dot A \rangle $ correspond to heat and work exchanges in quantum thermodynamics~\cite{binder2018thermodynamics,
QthermoDeffner2019}. For observables without explicit time dependence, one simply has $\dota = d\langle A \rangle/dt$.

An equation of motion analogous to~\eqref{eq:generaldyn} has been proven for classical stochastic systems~\cite{nicholson2020timeinformation} and is known as the Price equation in evolutionary biology~\cite{frank2020fundamental,
price1970selection}. 
Applying the Cauchy-Schwarz inequality, we can derive the following upper bound on the change in the expectation value due to state changes, $\dota \nobreak \coloneqq \nobreak \tr{A \tfrac{d\rho_t}{dt} } \nobreak = \nobreak \cov\left( \slog,A \right)$, 
which generalizes the Mandelstam and Tamm speed limit on observables: 
\begin{align}
\label{eq:generalbound}
\big| \dota \big| &=\left| \cov\left( \slog,A \right) \right| \leq \Delta A \Delta \slog = \Delta A \sqrt{\info }.
\end{align}
Here $\info \coloneqq (\Delta \slog)^2$ is the \emph{quantum Fisher information}. For a density matrix with a spectral decomposition $\rho_t = \sum_j p_j \ket{j}\!\bra{j}$, it 
is given by~\cite{BraunsteinCaves1994,
 braunstein1996generalized,
 paris2009quantum}
\begin{align}
\label{eq:qFisher}
\info = (\Delta \slog)^2 = 2 \sum_{jk}^d  \frac{\left| \bra{j} \dt \rho_t \ket{k}\right|^2}{p_j + p_k}.
\end{align}
The definition of $\info$ and the proof of Eq.~\eqref{eq:generalbound} assume differentiable dynamics and that $p_j \neq 0$  
$\forall j$~\footnote{
We show in Appendix~\ref{app:support} that, when states with $p_j = 0$ play a role in the evolution of an observable, there is a correction to Eq.~\eqref{eq:generalbound} that is upper bounded by $\left| \sum_{p_j = 0} \dot p_j \bra{j} A \ket{j} \right|$. This correction is typically small for most times in a differentiable evolution and null for states with constant support.}.
In order to ease notation, we omit explicit time-dependence of $\info$, $p_j$, $\{ \ket{j} \}$, $\slog$, and $A$.

Equation~\eqref{eq:generalbound} shows that the uncertainty of an observable and the quantum Fisher information limit the speed with which the mean of the observable evolves.
The Fisher information $\info$ originates from parameter estimation theory, where it
bounds the uncertainty with which a parameter---$t$ in this case---can be determined~\cite{helstrom1969quantum,
 holevo2011probabilistic,
 BraunsteinCaves1994,paris2009quantum,
sidhu_geometric_2020}.
Speed limits in Hilbert space~\cite{FrowisPRA2012,
 DavidovichPRL2013,
 toth2014quantum,
 PiresPRX2016,
 GessnerGeometricPRA2018} and speed limits for observables in closed systems~\cite{braunstein1996generalized,
 juliafarre2018bounds}
have also been linked to bounds from parameter estimation theory. 
In fact, when $A$ has no explicit time dependence, Eq.~\eqref{eq:generalbound} can be derived from the quantum Cram\'er-Rao bound by restricting to functions of $\langle A \rangle$ as (typically sub-optimal) estimators of the parameter $t$ (Appendix~\ref{app:ParamEst}).
Note, though, that the quantum Cram\'er-Rao bound encompasses arbitrary estimators and arbitrary measurements performed on the system. 
Thus, the generality of the Cram\'er-Rao bound  
leaves room for tighter bounds for particular estimators.
 We will exploit this in the next section to derive speed limits on $\langle A \rangle$ that are tighter than Eq.~\eqref{eq:generalbound} and than speed limits implied by Cramer-Rao bounds.

For a given speed $\dota$, the uncertainty bound~\eqref{eq:generalbound}, which we will refer to as the \emph{Cram\'er-Rao speed limit}, implies a direct trade-off between how certain an observable is and  
the Fisher information $\info$.
Roughly speaking, $\info$ is a measure of 
 stationarity in the system~\cite{flynn2014measuring,
 nichols2015order}:
  a small value implies a weak change of $\rho_t$ in time, which hinders the rate of change of any observable.
  Similarly, $\Delta A = 0$ implies a state $\rho_t$ supported on the subspace spanned by eigenvectors $\{ \ket{\lambda_a^l} \}$ corresponding to a single eigenvalue $\lambda_a$ of $A$. Then, the function $  \tr{\rho_t \sum_{l} \ket{\lambda_a^{l}}\!\!\bra{\lambda_a^{l}}} \nobreak = \nobreak \langle A\rangle / \lambda_a \nobreak \nobreak =\nobreak 1$ is at a maximum, implying a null rate of change, which translates into $\dota = 0$.
  On the other extreme, fast observable dynamics requires large fluctuations and large Fisher information.

Following
Mandelstam and Tamm, we identify the characteristic timescale $\tau_\fancya\nobreak\coloneqq\nobreak\Delta A/{\big| \dota \big|}$ over which the expectation value of an observable changes by a standard deviation~\cite{mandelstamtamm1945}.
Combining this definition with the inequality (\ref{eq:MT}), 
Mandelstam and Tamm established the time-energy uncertainty relation $\tau_A \Delta H \geq 1/2$, valid for isolated systems evolving with a constant Hamiltonian $H$ \cite{mandelstamtamm1945,BookMessiah}; see~\cite{booktimevol1,
Aharonov1990,
braunstein1996generalized} as well. 
For pure states $\info \nobreak=\nobreak4(\Delta H)^2$ and Mandelstam-Tamm's time-energy uncertainty can be re-expressed as $\tau_A \sqrt{\info }\nobreak\geq\nobreak1$.
Bound~\eqref{eq:generalbound} extends this to states 
 following arbitrary differentiable dynamics,
  \begin{align}
  \label{eq:time-information}
  \tau_A \sqrt{\info } \geq 1,
  \end{align}
and constitutes a time-information uncertainty relation
 that holds universally for quantum systems.

It is worth noting that uncertainty relations
are not purely a feature of quantum mechanics.
In fact, Ref.~\cite{nicholson2020timeinformation} shows that a bound analogous to Eq.~\eqref{eq:time-information} holds for classical stochastic systems. 
How, then, does the interplay of quantum and classical dynamics contribute to the speed at which an observable can evolve?

\section{Speed limits for coherent and incoherent dynamics}
\label{sec:coherent-incoherent}

The state $\rho_t$ of a quantum system evolving under arbitrary differentiable trace-preserving dynamics 
 can be written as
$\rho_t = U_t \, \chi_t \, U_t^\dag$. 
The unitary operator $U_t$ connects the time-dependent eigenbasis of $\rho_t$ to the eigenbasis $\{ \ket{j}_0\}$ of the initial state by $\ket{j}\nobreak=\nobreak U_t \ket{j}_0$, and $\chi_t\coloneqq \sum_j p_j  \ket{j}_{0}   \prescript{}{0\!}{\bra{j}} $  is a diagonal density matrix with the same eigenvalues as $\rho_t$. 
Defining the Hermitian operator 
$H_t \coloneqq i \tfrac{d U_t}{dt} U_t^\dag$ as the `Hamiltonian',
one can express the evolution of $\rho_t$ as~\cite{avron1987adiabatic,
GirolamiPRL2019,Alipour20}
\begin{align}
\label{eq:dynamics}
\dt \rho_t = - i[H_t,\rho_t] + U_t \frac{d \chi_t}{dt} U_t^\dag.
\end{align}
This equation describes the dynamics of any quantum system 
 with a continuous physical evolution where probability is conserved, including  non-Markovian 
 dynamics~\cite{li2018Markov} and non-linear dynamics stemming from continuous monitoring~\cite{Jacobs_2006} or balanced gain and loss~\cite{Alipour20}. 
 In the widely relevant case when the system obeys a Lindblad equation for Lindblad operators $\Gamma_\alpha$ with rates $\gamma_\alpha$ that cause incoherent dynamics, 
one has $U_t \tfrac{d \chi_t}{dt} U_t^\dag\nobreak= \nobreak\sum_\alpha \gamma_\alpha \left( \Gamma_\alpha \rho  \Gamma_\alpha^\dag - \tfrac{1}{2} \{ \Gamma_\alpha^\dag \Gamma_\alpha, \rho  \} \right)$.
 (Note that 
 Lindblad operators can also induce unitary dynamics, in which case they would contribute to $H_t$~\cite{LukaszPC,Lukaszpaper}.)

The first term in Eq.~\eqref{eq:dynamics} represents  \emph{coherent}, entropy-preserving  evolution. The second, \emph{incoherent}, term corresponds to changes in the state's eigenvalues and therefore in the entropy of the system. 
Note that the case with no coherent contribution $H_t=0$ gives rise to a description equivalent to a probability distribution $\{ p_j \}$ following classical stochastic dynamics~\cite{Seifert_2012}.  In contrast, even if coherence is not a uniquely quantum trait~\cite{lostaglio2020certifying}, coherent dynamics due to $H_t \neq 0$ is characteristic of quantum systems.
 In this sense, 
 one could broadly identify the coherent and incoherent terms to correspond to `quantum' and `classical' contributions to the evolution, respectively~\cite{PiresPRX2016,GirolamiPRL2019}. This identification is further supported by the fact that, if the eigenstates $\ket{j}$ involved in purely incoherent dynamics are classical (e.g., bit strings representing product states of computational basis states of two-level systems), the resulting incoherent dynamics is indeed classical. On the other hand, one should take the `classical' label for incoherent dynamics with a grain of salt since eigenstates $\ket{j}$ can be highly nontrivial, in which case incoherent dynamics can also be quantum, for instance driving product states into entangled ones~\footnote{Consider two qubits in a state with a spectral decomposition $\rho = p_\textnormal{ent} \ket{\Psi_\textnormal{ent}}\!\bra{\Psi_\textnormal{ent}} \nobreak + \nobreak p_\textnormal{prod} \ket{\Psi_\textnormal{prod}}\!\bra{\Psi_\textnormal{prod}}$, where $\ket{\Psi_\textnormal{ent}} = \tfrac{1}{\sqrt{2}}\left( \ket{00} + \ket{11} \right)$ and $\ket{\Psi_\textnormal{prod}} =  \ket{01}$. A change in the probabilities $\{ p_\textnormal{ent}, p_\textnormal{prod}\}$ has a description in terms of incoherent dynamics even though it involves a very quantum process where entanglement between the qubits can grow.}.

To extend the coherent-incoherent separation to the observable of interest $A\nobreak\coloneqq\nobreak A_C\nobreak+\nobreak A_I$, we define the relevant components to the evolution of the observable for a system with purely coherent ($\dot \chi_t = 0$) or purely incoherent ($H_t = 0$) 
dynamics, 
\begin{align}
\label{eq:classandquantcomponents}
A_C \coloneqq \sum_{j\neq k}^d A_{jk} \ket{j}\!\bra{k} \quad \textnormal{and} \quad A_I \coloneqq \sum_{j}^d A_{jj} \ket{j}\!\bra{j}.
\end{align}
In this separation, we take the time-dependent basis $\{ \ket{j} \}$ that diagonalizes state $\rho_t$. The coherent-incoherent division is thus relative to the state of the system.

As a final ingredient to our construction, we define Hermitian operators $\slogq$ and $\slogc$, with $\slog = \slogq + \slogc$, that will characterize the speed due to coherent and incoherent dynamics:
\begin{subequations}
\label{eq:saturationV2}
\begin{align}
\label{eq:saturationV2quant}
\slogq &\coloneqq -2i\sum_{j \neq k}^d    \frac{\bra{j}\nobreak [H_t,\rho_t] \ket{k}}{(p_j\nobreak+\nobreak p_k)} \ket{j} \! \bra{k}, \\
\label{eq:saturationV2class}
\slogc &\coloneqq \sum_{j}^d  \frac{d \ln p_j}{dt} \ket{j} \! \bra{j}.
\end{align}
\end{subequations}
With these operators, one can 
separate the quantum Fisher information  
 into coherent and incoherent contributions $\info \nobreak =\nobreak \info^C \nobreak+\nobreak \info^I$ (Appendix~\ref{app:cohincoh}):
\begin{subequations}
\label{eq:Fisher}
\begin{align}
\info^C &\coloneqq \left( \Delta \slogq \right)^2 = 2 \sum_{j \neq k}^d  \frac{ \Big|\! \bra{j} [H_t,\rho_t] \ket{k} \!\Big|^2 }{p_j + p_k}, \\
\info^I &\coloneqq \left( \Delta \slogc \right)^2 = \sum_j^d p_j \left( \frac{d}{dt} \ln p_j \right)^2.
\end{align}
\end{subequations}
Note that $\info^I$ is the classical Fisher information of the probability distribution $\{ p_j  \}$ 
~\cite{Fisher1922,KimEntropy2018}, 
 which also admits an interpretation in terms of the variance in the \emph{surprisal rate} $\{(\slogc)_j = -\frac{d}{dt}\ln p_j\}$ associated with the eigenvalue distribution $\{p_j\}$~\cite{VedralReview2002}. Meanwhile, $\info^C$ is the quantum Fisher information for a system evolving unitarily. For pure states, $\info^C = 4 (\Delta H_t)^2$~\cite{BraunsteinCaves1994}.

This construction, which separates the change of an observable $A$ into coherent and incoherent contributions, 
allows for the derivation of bounds that are tighter than the Cram\'er-Rao speed limit~\eqref{eq:generalbound}.
We prove in Appendix~\ref{app:cohincoh} that, for differentiable dynamics, $\dota_C \nobreak \coloneqq \nobreak  \tr{A_C \frac{d\rho_t}{dt}} \nobreak = \nobreak \cov\left(A_C,\slogq \right)$ and $\dota_I \nobreak \coloneqq \nobreak  \tr{A_I \frac{d\rho_t}{dt}} \nobreak = \nobreak \cov\left(A_I,\slogc \right)$. Therefore, 
\begin{subequations}
\label{eq:bound2} 
\begin{align}
\label{eq:bound2q}
 \big|  \dota_C \big| &= \left| \cov\left(A_C,\slogq \right) \right| \leq \Delta A_C \, \sqrt{\info^{C}} , \\
\label{eq:bound2c}
  \big| \dota_I \big| &= \big| \cov\left(A_I,\slogc \right) \big|   \leq \Delta A_{I} \, \sqrt{\info^{I}}
 \end{align} 
 \end{subequations}
set bounds on the rate of change of an observable that isolate the role played by coherent and incoherent dynamics.
Equation~\eqref{eq:bound2c} is identical to the one derived in~\cite{nicholson2020timeinformation} for classical stochastic systems, where $-(\slogc)_j$ correspond to the \emph{surprisal rates}.
This coherent-incoherent decomposition and the corresponding bounds constitute one of the main results of the paper.

These results provide a foundation for a number of speed limits on observables.
The reverse triangle inequality~\cite{Abramowitz} yields a lower bound on the speed $ \dota\nobreak = \nobreak\tr{A \frac{d\rho_t}{dt}}=\nobreak\dota_C\nobreak+\nobreak\dota_I$, i.e., a \emph{coherent-incoherent lower speed limit} 
\begin{align}
\label{eq:boundlower} 
\!\!\! \big|  \dota \big| &\geq 
\max  \left\{ \big|  \dota_C \big| - \Delta A_{I} \sqrt{\info^{I}}  
 \, , \, \big|  \dota_I \big| - \Delta A_{C} \sqrt{\info^{C}} \right\},
\end{align}
made possible by the division of the system dynamics into two contributions. 
To the best of our knowledge, this is the first derivation of general lower speed limits on the evolution of open quantum systems.

Equation~\eqref{eq:bound2} also implies a new upper bound 
 singling out additive contributions from the uncertainties of the coherent and incoherent parts $A_{C}$ and $A_{I}$ of the observable, 
\begin{align}
\label{eq:bound2v2} 
\big| \dota \big| &\leq \min  \left\{ \big|  \dota_C \big| + \Delta A_{I} \sqrt{\info^{I}}  
 \, , \, \big|  \dota_I \big| + \Delta A_{C} \sqrt{\info^{C}} \right\} \nonumber \\
&\leq \Delta A_C \, \sqrt{\info^{C}} + \Delta A_{I} \, \sqrt{\info^{I}}.
\end{align}
This bound limits the speed of systems following coherent quantum dynamics, as well as systems following incoherent classical dynamics;  a \emph{coherent-incoherent speed limit}.
While the Cram\'er-Rao speed limit~\eqref{eq:generalbound}
can be saturated by properly chosen observables, we show in Section~\ref{sec:saturation} below that the
coherent-incoherent
speed limit~\eqref{eq:bound2v2} is tighter for any $A$, imposing stricter constraints on the speed of evolution.

Two limiting cases demonstrate the scope of the upper bound. 
For purely coherent evolution, constant probabilities $\dot p_j(t) = 0$ imply $\info^C = 0$, in which case the
coherent-incoherent 
speed limit recovers the Mandelstam-Tamm bound for observables, generalized to allow for an explicit time-dependence in the Hamiltonian. 
In the opposite extreme of purely incoherent dynamics, $H_t = 0$, it recovers the classical speed limit recently derived in~\cite{nicholson2020timeinformation} for classical stochastic dynamics.

More generally, consider a system in state $\rho_t$ with dynamics that results in $\info^I$ and $\info^C$.
The coherent-incoherent speed limits
tell us that
some observables evolve following incoherent dynamics characterized by $\info^I$,  while others follow the coherent dynamics with $\info^C$. 
The evolution of
observables with $\Delta A_C \nobreak = \nobreak \sqrt{\sum_{j \neq k} p_j |A_{jk}|^2 }=0$, which commute with the state, is guided by the incoherent contribution to the evolution.
Meanwhile, observables with $\Delta A_I = 0$ are driven solely by the Hamiltonian.

The coherent-incoherent split of an observable motivates the definition of characteristic timescales $\tau_{A_C} \nobreak\coloneqq\nobreak\Delta A_C /{\big| \dota_C \big|}$ and $\tau_{A_I}\nobreak\coloneqq\nobreak\Delta A_I/{\big| \dota_I \big|}$ for $A_C$ and $A_I$, respectively, over which each of them change by their standard deviations.
From Eq.~\eqref{eq:bound2}, their time-information uncertainty bounds are  
\begin{align}
\label{eq:time-information2}
  \tau_{A_C} \sqrt{\info^C} \geq 1 
  \qquad  \textnormal{and} \qquad
  \tau_{A_I} \sqrt{\info^I} \geq 1.
\end{align}
The first bound generalizes Mandelstam-Tamm's time-energy uncertainty relation to the coherent component of the evolution. The second bound generalizes the classical time-information uncertainty bound from~\cite{nicholson2020timeinformation} to the incoherent contribution to the dynamics of an observable for a quantum system with arbitrary differentiable dynamics. 
Note, though, that these apply to  general regimes in which an open quantum system evolves under both coherent and incoherent dynamics. 
They provide alternate uncertainty relations to the quantum time-information uncertainty bound~\eqref{eq:time-information} that single out coherent and incoherent characteristic timescales.

\section{Speed limits in terms of energy variances}
\label{sec:boundsfisher}

One may be interested in bounds that depend on physical quantities that are more accessible than the Fisher informations central to our results above.   
It turns out that the following holds: 
\begin{align}
\label{eq:fisherbounds}
\info^C \leq 4 (\Delta H_t)^2 \qquad \textnormal{and} \qquad \info^I \leq 4 (\Delta H_t^{\textnormal{int}})^2.
\end{align}
Here we assume that the source of non-unitary dynamics is entanglement with an environment via a Hamiltonian $H_t^{\textnormal{int}}$  that includes all terms with support on both the system and the environment. The standard deviation $\Delta H_t^{\textnormal{int}}$  is calculated in the joint state of the system and the environment. The bound in Eq.~(\ref{eq:fisherbounds}) on $\info^C$ was proven by Braunstein and Caves in~\cite{braunstein1996generalized}. We prove the new (loose) bound in Eq.~(\ref{eq:fisherbounds}) on $\info^I$ in Appendix~\ref{app:FisherInformtions}. When combined with the coherent-incoherent speed limits~\eqref{eq:bound2v2} and~\eqref{eq:boundlower}, Eq.~\eqref{eq:fisherbounds} implies upper and lower bounds on $| \dota |$ in terms of energy uncertainties.

Certain physical quantities naturally evolve solely under incoherent dynamics.
The von Neumann entropy $S\nobreak\coloneqq\nobreak -\tr{\rho_t  \ln \rho_t }$ of a system is one example. Using $\dot S\nobreak =\nobreak - \tr{d\rho_t  /dt  \ln \rho_t }$~\cite{Das2018} and choosing $A_I\nobreak =\nobreak - \ln \rho_t$ in Eq.~\eqref{eq:bound2c} gives a bound   
\begin{align}
\label{eq:boundentropy}
\big| \dot S \big| \leq \Delta S \, \sqrt{\info^I} \leq 2 \, \Delta S \, \Delta H_t^{\textnormal{int}},
\end{align}
set by the variance $\left( \Delta S \right)^2 \coloneqq \tr{\rho_t (\ln \rho_t )^2} - S^2$ of the `surprisal operator' $[-\ln \rho_t]$, whose expectation value is the von Neumann entropy. 
 That is, an uncertainty relation bounds the change in entropy of any open quantum system in terms of fluctuations in energy
and in surprisal $\left[-\ln \rho_t\right]$
 ($\Delta S$ also plays the interesting role of determining the possible transitions between states of a quantum system~\cite{ HenrikarXiv2020}). 
 Using that $\Delta S \leq \sqrt{ (\ln(d-1))^2/4 + 1}$~\cite{ReebIEEE2015,
HenrikarXiv2020} and that $\Delta H_t^{\textnormal{int}} \leq \| H_t^{\textnormal{int}} \|$, where $\| H_t^{\textnormal{int}} \|$ is the operator norm, we recover  
the {\it Small Incremental Entangling theorem}~\cite{BravyiPRA2007,
MarienCommMathPhys2016}, $\big| \dot S \big| \lesssim \ln d \| H_t^{\textnormal{int}} \|$, in the case of no ancillary systems (Appendix~\ref{app:SIE}). 

Even though the Hamiltonian is generally not of the form of $A_C$, since it need not commute with $\rho_t$, a direct calculation with Eq.~\eqref{eq:dynamics} shows that the energy of the system changes solely due to the time-dependence of $H_t$ or under the incoherent contribution to the dynamics. Thus, from Eqs.~\eqref{eq:bound2c} and~\eqref{eq:fisherbounds} it holds that 
\begin{align}
\label{eq:boundenergy}
 \left| \tr{ \frac{d \rho_t}{dt} H_t} \right| \leq \Delta H_C \, \sqrt{\info^C} \leq \Delta H_t \, \Delta H_t^\textnormal{int},
\end{align}   
where $H_C$ is the diagonal component of $H_t$ as defined in Eq.~\eqref{eq:classandquantcomponents} and we used that $\Delta H_C \leq \Delta H_t$.
This result sets bounds on what is typically identified as the heat flux in the quantum thermodynamics of open systems~\cite{binder2018thermodynamics,
QthermoDeffner2019}~\footnote{See~\cite{Sahar2019} for a critical analysis of the interpretation of heat and work in open quantum systems.}. 
This new speed limit
 is a quantum analog of the bound derived in~\cite{nicholson2020timeinformation} for classical stochastic thermodynamics.

\section{Saturation and tightness of the speed limits}
\label{sec:saturation}
Here, we focus on the tightness of the main new coherent-incoherent speed limit~\eqref{eq:bound2v2} and how it compares to the Cram\'er-Rao speed limit~\eqref{eq:generalbound}. We find that the former provides a strictly tighter constraint on the rate of change of observables than the speed limit implied by the quantum Cram\'er-Rao bound. To see this, we use  $\info = \info^C+\info^I$ and  $(\Delta A)^2\nobreak =\nobreak (\Delta A_C)^2\nobreak+ \nobreak (\Delta A_I)^2$ to prove that (Appendix~\ref{app:comparison})
\begin{align}
\label{eq:saturation}
\tfrac{ \Delta A \sqrt{\info}  }{ \Delta A_C \sqrt{\info^C} + \Delta A_I \sqrt{\info^I} } = \sqrt{1+\tfrac{\left( \Delta A_C \sqrt{\info^I} - \Delta A_I \sqrt{\info^C} \right)^2}{\left( \Delta A_C \sqrt{\info^C} + \Delta A_I \sqrt{\info^I} \right)^2}},
\end{align}
which quantifies the improvement that the coherent-incoherent speed limit provides. The two bounds coincide when $\Delta A_C \sqrt{\info^I}\nobreak =\nobreak \Delta A_I \sqrt{\info^C}$. This condition occurs for pure states $\rho_t = \ket{k}\!\bra{k}$ with purely coherent dynamics [$\left(\Delta A_I\right)^2 \nobreak  = \nobreak  {\sum_j p_j A_{jj}^2 - (\sum_j p_j A_{jj})^2} \nobreak = \nobreak 0$  and $\info^I \nobreak =\nobreak  0$], in which case bounds~\eqref{eq:bound2v2} and~\eqref{eq:generalbound} coincide with the one by Mandelstam and Tamm~\cite{mandelstamtamm1945}. It also occurs for diagonal observables with purely incoherent dynamics ($\Delta A_C \nobreak = \nobreak 0$ and $\info^C \nobreak=\nobreak 0$), in which case both bounds coincide with the classical one derived in~\cite{nicholson2020timeinformation}. In contrast, whenever $\Delta A_C \sqrt{\info^I}\nobreak \neq\nobreak \Delta A_I \sqrt{\info^C}$, our new coherent-incoherent speed limit~\eqref{eq:bound2v2} is tighter.

 Taking advantage of the fact that $\dota \nobreak=\nobreak \cov(A,\slog)$, $\dota_C\nobreak = \nobreak\cov(A_I,\slogq)$, and $\dota_I \nobreak=\nobreak \cov(A_I,\slogc)$, one can identify observables that evolve at the limiting speeds. The Cram\'er-Rao speed limit~\eqref{eq:generalbound} saturates when $\cov(A,\slog)\nobreak =\nobreak \Delta A \, \Delta \slog$, which is the case for observables $A \propto \slog$.  Similarly, observables such that $A_C \propto \slogq$ and $A_I \propto \slogc$ saturate the coherent and incoherent speed limits~\eqref{eq:bound2}~\footnote{Note that $A_C \propto \slogq$ and $A_I \propto \slogc$ are sufficient but not necessary conditions to saturate the coherent and incoherent speed limits~\eqref{eq:bound2}. For example $A_C \propto \slogq \nobreak+\nobreak \kappa_C \id$ and $A_I \propto \slogc \nobreak+\nobreak \kappa_I \id$ also saturate them for constants $\kappa_C$ and $\kappa_I$, given that additive constants do not change covariances.}.  This property defines the role of Hermitian operators $L_C$ and $L_I$, evaluated at $\rho_t$, as 
  observables that evolve at their speed limits.
   We can also see how saturation of
the coherent-incoherent speed limit can occur in cases when the Cram\'er-Rao speed limit is looser: an operator $A \nobreak=\nobreak \alpha_C \slogq \nobreak+ \nobreak\alpha_I \slogc$ saturates the former but not the latter except when $\alpha_C \nobreak=\nobreak \alpha_I$.
 
In fact, the `\emph{fast}' coherent and incoherent operators $L_C$ and $L_Q$ are orthogonal to each other with respect to the inner product defined by the symmetrized covariance, $\cov (\slogq, \slogc) = 0$. 
Thus, they 
form part of an orthogonal \emph{basis of speed operators} $\{\slogq , \slogc , \slog_0^{1}, \dots , \slog_0^{d^2-2}\}$ that spans the space of Hermitian operators. 
Since evolution occurs only under coherent or incoherent dynamics, and $\cov (\slogq, \slog_0^{n})\nobreak =\nobreak \cov (\slogc, \slog_0^{n}) \nobreak =\nobreak 0$ by construction, the `\emph{still}' operators $\{ \slog_0^n \}$ do not evolve.
Then, any observable expressed in terms of the preferred basis as $A\nobreak =\nobreak \alpha_C \slogq \nobreak + \nobreak \alpha_I \slogc \nobreak +\nobreak  \sum_n \alpha_n \slog_0^n$ 
 evolves with a speed $\dota = \alpha_C  {\info^C} + \alpha_I {\info^I}$.

We can use this construction to understand the situations in which the reverse triangle inequality used to derive the lower speed limit~\eqref{eq:boundlower} saturates. The bound $\big| \dota \big| \nobreak\geq\nobreak \big| \dota_C \big| \nobreak-\nobreak \Delta A_{I} \sqrt{\info^{I}} $ is saturated if 
$A_I = - \alpha_I \slogc$ with $\alpha_I \ge 0$, 
 while $\big| \dota \big| \nobreak\geq\nobreak \big|  \dota_I \big| \nobreak-\nobreak \Delta A_{C} \sqrt{\info^{C}} $ saturates for 
 $A_C = - \alpha_C \slogq$ with $\alpha_C \ge 0$. 
Once again, the preferred operator basis that contains $\slogq$ and $\slogc$ serves to characterize the speed of an observable and how far it is from saturating the upper and lower speed limits.

\begin{figure*}
\centering
\includegraphics[width=\textwidth]{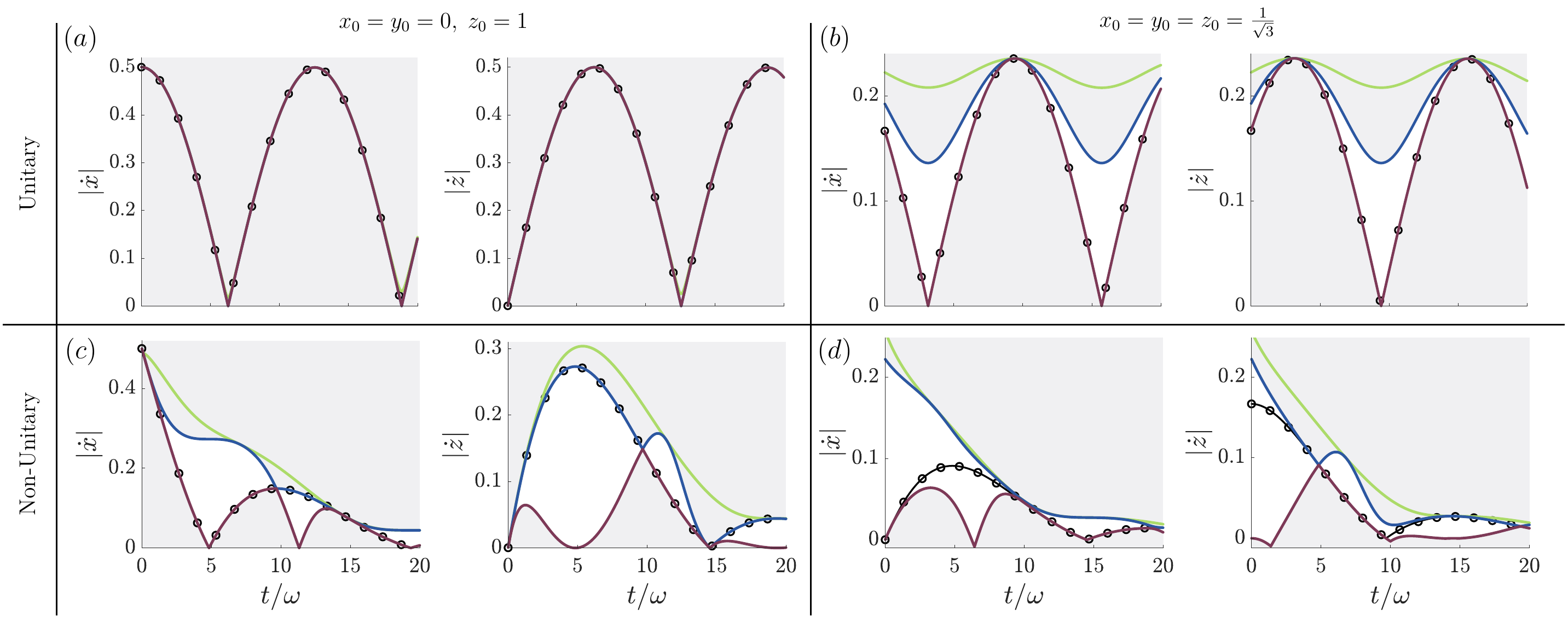}
\textcolor{black}{\contour{black}{$\multimap$--}}   \small speed  \quad \textcolor{S_Green}{\contour{S_Green}{---}}  \small Cram\'er-Rao speed limit
\quad \textcolor{Blue}{\contour{Blue}{---}}   \small coherent-incoherent speed limit
\quad
   \textcolor{S_Brown}{\contour{S_Brown}{---}}  \small coherent-incoherent lower speed limit
 \caption{
\label{fig:fig2}
\textbf{Speed limits for  observables on a qubit.}
The speed limit~\eqref{eq:generalbound}, derivable from the Cram\'er-Rao bound, and the coherent-incoherent speed limits~\eqref{eq:bound2v2} and~\eqref{eq:boundlower} impose constraints on the rate $\dota$ at which observables 
evolve. 
We illustrate this on a qubit with a state parametrized by $\rho \nobreak = \tfrac{1}{2}\nobreak(\id \nobreak+\nobreak x \sigma_x \nobreak+\nobreak y \sigma_y \nobreak+\nobreak z \sigma_z) $ and driven by the Hamiltonian $H = (\omega/2) \sigma_y$ and dephasing along $\sigma_z$ with a rate $\kappa$. 
Black circled lines denote the  speeds $|\dot x|$ and $|\dot z|$ of observables $\sigma_x$ and $\sigma_z$, 
and blue (red) lines denote the coherent-incoherent (lower) speed limits.  The grayed areas denote rates forbidden by the coherent-incoherent speed limits. 
In green, the looser Cram\'er-Rao speed limit  appears inside the forbidden region.
(a,c) The qubit is initialized in state $z_0=1$, and thus evolves in the $y=0$ plane on and inside the Bloch sphere, which in turn implies that both observables $\sigma_z$ and $\sigma_x$ have coherent-incoherent decompositions that satisfy conditions for saturation of Eq.~\eqref{eq:bound2}, with 
 $\sigma_{\{z,x\}} = \alpha_C^{\{z,x\}} \slogq + \alpha_I^{\{z,x\}} \slogc$.  
For unitary dynamics [$\kappa = 0$, (a)], this also means that 
the Cram\'er-Rao speed limit saturates and the three curves coincide. 
 For open-system dynamics [$\kappa \neq 0$, (c)] the red and black curves coincide, but the Cram\'er-Rao speed limit
 is looser except when $\alpha_C^{\{z,x\}} =\alpha_I^{\{z,x\}}$. 
Alternatively, whenever  
$\alpha_C^{\{z,x\}}$ and $\alpha_I^{\{z,x\}}$  have the same (different) sign,
 the coherent-incoherent (lower) speed limits saturate.
(b,d)
The coherent-incoherent speed limits~\eqref{eq:bound2v2} and~\eqref{eq:boundlower} are not tight for a qubit initialized in $x_0 = y_0 = z_0 = 1/\sqrt{3}$, when the state does not evolve within the $y=0$ plane and the observables no longer have a decomposition solely in terms of $\slogq$ and $\slogc$. Nevertheless, these bounds serve to constrain observables' dynamics more than the Cram\'er-Rao speed limit.
}
\end{figure*}

The following example illustrates how to construct observables that saturate the coherent and incoherent bounds. 
Consider a qubit with a Hamiltonian $H\nobreak =\nobreak \tfrac{\omega}{2} \sigma_y \nobreak$
and with incoherent dynamics driven by dephasing along $\sigma_z$ with a rate $\kappa$, described by $U_t \tfrac{d \chi_t}{dt} U_t^\dag = - \kappa [\sigma_z,[\sigma_z,\rho_t]]$ in Eq.~\eqref{eq:dynamics}.
Let us consider the case when the qubit starts with $y = \tr{\rho_t \sigma_y} = 0$ (Fig.~\ref{fig:fig2}, left column).
The state of the qubit is parametrized as $\rho_t = \frac{\id + x \sigma_x + z \sigma_z}{2}$, where $x = \tr{\rho_t \sigma_x}$ and $z = \tr{\rho_t \sigma_z}$ are real numbers such that $x^2 + z^2 \leq 1$. 
Since the generators of dynamics preserve $y=0$, we have that $\cov(\slogq,\sigma_y) = \cov(\slogc,\sigma_y) =0 $ (Appendix~\ref{app:saturation}). 
Then, $\{ \slogq, \slogc, \sigma_y, \id\}$ forms a complete basis of Hermitian operators and,
since observables $\sigma_x$ and $\sigma_y$ are orthogonal to $\id$ and $\sigma_y$, they can be written as
 $\sigma_{\{z,x\}} \nobreak = \nobreak \alpha_C^{\{z,x\}} \slogq + \alpha_I^{\{z,x\}} \slogc$. 
 This means that the coherent and incoherent terms of observables $\sigma_x$ and $\sigma_z$ evolve at the speed limits allowed by their respective 
  bounds~\eqref{eq:bound2}. For the case of unitary dynamics ($\kappa = 0$),  this also means that the Cram\'er-Rao speed limit~\eqref{eq:generalbound}
  saturates. However, when $\kappa \neq 0$, the latter is loose except when $\alpha_Q^{\{z,x\}} = \alpha_C^{\{z,x\}}$. Finally, saturation of the coherent-incoherent speed limits~\eqref{eq:bound2v2} and~\eqref{eq:boundlower}  depends on the relative signs of $\alpha_C^{\{z,x\}} $ and $ \alpha_I^{\{z,x\}}$. 
  Instead, if the system starts with $y \neq 0$, the fast operators have components on $\{\sigma_x,\sigma_y,\sigma_z\}$, which means that observables $\sigma_{\{x,y,z\}}$ do not have expressions solely in terms of the fast operators $\slogq$ and $\slogc$. Thus, the coherent-incoherent speed limits are typically not saturated in this case (Fig.~\ref{fig:fig2}, right column).

These observable-dependent bounds can vary significantly from one observable to another for a given system, as shown by the bounds on the speeds of $\sigma_x$ and $\sigma_z$ that Fig.~\ref{fig:fig2} depicts. This example illustrates the spirit of our initial aim: to find bounds that better capture the speed of physical observables than speed limits in Hilbert space. The bounds obtained from the coherent-incoherent decomposition of the dynamics capture the dynamics better than those derivable from the quantum Cram\'er-Rao bound.

\section{Bounds on integrated quantities}
\label{sec:geometry}
 The Fisher information has interesting connections to the geometry of the space of probability distributions and of state space~\cite{DavidovichPRL2013,WoottersPRD1981,
 Aharonov1990,BraunsteinCaves1994,
BookBengtssonGeometry2006,
CrooksPRL2007,GessnerGeometricPRA2018,
NicholsonPRE2018}.
For small changes in $t$, 
the quantum Fisher information is  
related to the Bures distance $D_B(\rho_t,\rho(t+\delta t))$ between neighboring states
by 
\begin{align}
\label{eq:buresmetric}
ds^2 = D_B^2(\rho_t,\rho_{t+dt}) = \frac{\info}{4} dt^2, 
\end{align}
defining a metric in the space of density operators~\cite{BookBengtssonGeometry2006}.
The Bures distance 
between any two states $\rho_1$ and $\rho_2$ 
is defined by $D_B(\rho_1,\rho_2)\nobreak\coloneqq\nobreak\sqrt{2}\sqrt{1- \sqrt{F(\rho_1,\rho_2)}}$, where $F(\rho_1,\rho_2)\nobreak\coloneqq\nobreak\left( \tr{\sqrt{\sqrt{\rho_2}\rho_1\sqrt{\rho_2}}} \right)^2$ is the Uhlmann fidelity. 
This connection has been exploited in the literature to derive a lower bound on the time needed for a system to evolve between orthogonal states~\cite{DavidovichPRL2013, [{We show in Appendix~\ref{app:qslHilbert} how to use the coherent-incoherent upper bound on observables to derive a tighter speed limit on the fidelity $F(\rho_t,\rho_0)$ between a pure initial state and the evolved state $\rho_t$}] foonote1}.

Equations~\eqref{eq:generalbound} and~\eqref{eq:buresmetric} lead to an integrated bound that relates the change in the observable to the Bures length $\mathcal{L}(\rho_i,\rho_f)$ of the path followed by the system:
\begin{align}
\label{eq:lengthconstraint}
\!\!\!\! \int_{0}^{\tau} \frac{\big|  \dota  \big|}{\Delta A}   dt \leq  \int_{0}^{\tau} \sqrt{\info} dt =   2  \int_{\rho_i}^{\rho_f} ds =  2\mathcal{L}(\rho_i,\rho_f) .
\end{align}
This new bound  
shows how the path taken by the system in state space puts
 constraints on the corresponding evolution of system observables. 
In particular, note that $A$ is not the Hamiltonian or the generator of evolution, but an arbitrary observable of interest. 
 The geodesic is the path that minimizes $\mathcal{L}(\rho_i,\rho_f)$
and its length
is given by the Bures angle length $\mathcal{L}_{\min}(\rho_i,\rho_f)\nobreak=\nobreak\arccos \left[ \sqrt{F(\rho_i,\rho_f)}\right]$~\cite{BookBengtssonGeometry2006,
PiresPRX2016}.
\begin{figure} 
  \centering  \includegraphics[trim=00 00 00 00,width=0.45 \textwidth]{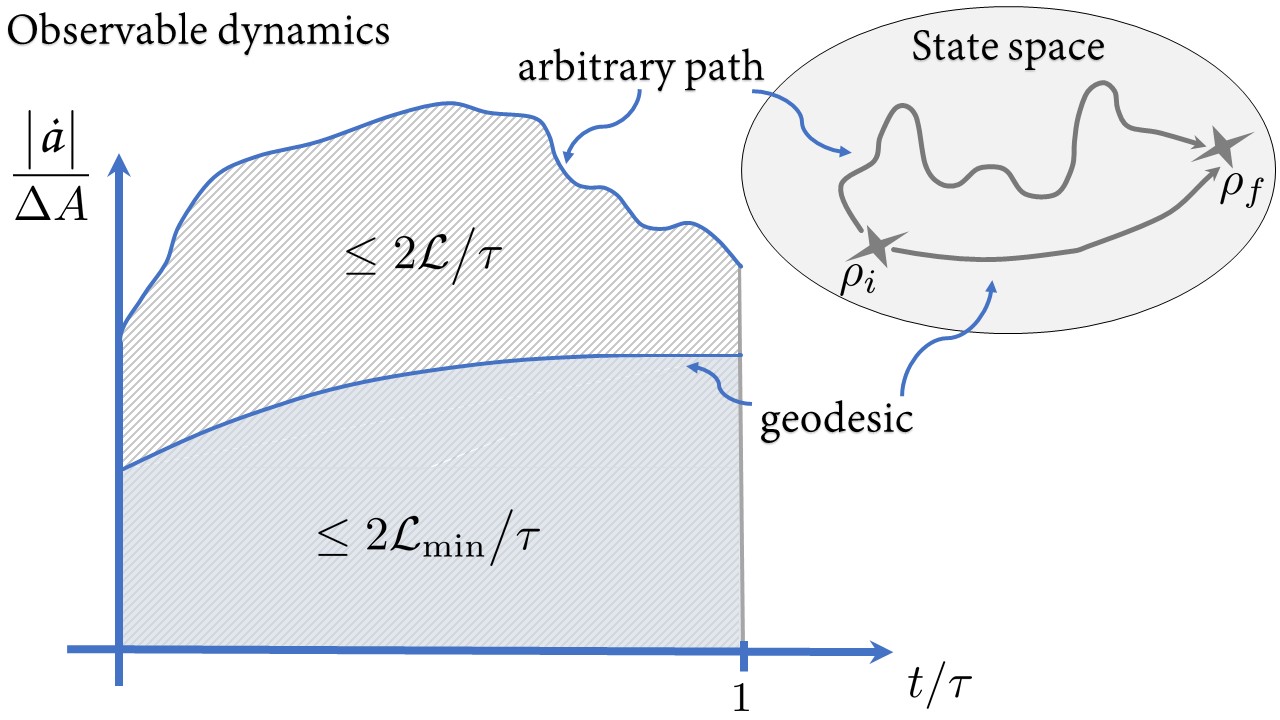} 
\caption{\label{fig:fig1}
\textbf{Constraints on evolution of observables.}
The integrated speed limits~\eqref{eq:lengthconstraint} and~\eqref{eq:integrated} place constraints on the rate of change of an observable, which depend on the path taken by the state of the system in state space.
The former bound says that the area under the curve $\big| \dota \big|/\Delta A$ of an arbitrary path is upper bounded by the path length $\mathcal{L}$, as $\int  \big|  \dota  \big|/\Delta A dt \leq  2\mathcal{L} $. 
Geodesics put a more stringent constraint on the dynamics of an observable: the area under the curve $\big| \dota \big|/\Delta A$ must satisfy $\int  \big|  \dota  \big|/\Delta A dt \leq  2\mathcal{L}_{\min} $. Note that $A$ need not be related to the Hamiltonian or the generator of the dynamics, but is rather an arbitrary observable of interest.
}
\end{figure}
Figure~\ref{fig:fig1}
illustrates the constraints that Eq.~\eqref{eq:lengthconstraint} imposes on different paths taken by the system in state space.

Along the lines similar to the previous integrated bound,
 the change of an observable due to the change in the state from $\rho_i$ to $\rho_f$ over a time $\tau$ satisfies (Appendix~\ref{app:Divergence})
\begin{align}
\label{eq:integrated}
\big| \fancya(\tau)   \big| &= \left| \int_{0}^{\tau} \dota \, dt \right| \leq 2 \int_{\rho_i}^{\rho_f} \Delta A \, ds  \\
& \leq 2 \sqrt{ \mathcal{J}(\rho_i,\rho_f) \frac{1}{\tau}  \int_{0}^{\tau}  \left( \Delta A\right)^2  dt. } \nonumber
\end{align}
The total change is thus bounded by the integrated fluctuations of the observable over the path that the state follows in Hilbert space.
This, in turn, is bounded by the integrated observable fluctuations and the \emph{divergence of the path} $\mathcal{J}(\rho_i,\rho_f) \nobreak \coloneqq \nobreak \tau \int_0^\tau \info dt$ in state space~\cite{CrooksPRL2007,NicholsonPRE2018,GirolamiPRL2019}.

The decomposition of dynamics into incoherent and coherent terms also provides a way to quantify the integrated contribution of each of them to the change in an observable. 
Often, quantum phenomena necessitate unitary dynamics in order to benefit from quantum mechanical advantages. For example, unavoidable incoherent effects stemming from experimental imperfections in isolating, preparing, or driving a system amount to errors in the resulting dynamics, hindering  quantum cryptographic protocols~\cite{RevModPhysQCrypto2002} and computing~\cite{NatureMonroe2016,
Naturezhang2017observation,
NatureQSup2019} or destroying quantum correlations~\cite{Bell1,Bell2,Bell3}.
Integrating bound~\eqref{eq:bound2} on  the incoherent contribution to the change of an observable gives a bound on how much the incoherent dynamics affects an observable:
\begin{align}
\label{eq:totalchangeincoh}
\big| \fancya(\tau) - \fancya_C(\tau) \big| = \big| \fancya_I(\tau) \big|  &= \left| \int_0^\tau \! \! \cov(A_I,\slogc) dt \right|\nonumber \\
&\leq \int_{0}^{\tau} \! \Delta A_I \sqrt{\info^I} dt.
\end{align} 
This bound can 
be interpreted as quantifying the error that incoherent dynamics induces to the desired change $\fancya_C(\tau)$ in an observable $A$, had the system evolved according to an ideal unitary evolution.
An observable that cannot discriminate  eigenstates of $\rho_t$ along its evolution, in the sense of $\bra{j} \! A \! \ket{j} \nobreak=\nobreak \bra{k}\! A\! \ket{k}$ $\forall j,k$, obeys $\Delta A_I \nobreak =\nobreak 0$ and thus does not accumulate an error due to incoherent evolution. 

Alternatively,  
\begin{align}
\label{eq:totalchangecoh}
\!\big| \fancya(\tau)\! -\! \fancya_I(\tau) \big|  &= \left| \int_0^\tau \cov(A_C,\slogq) dt \right| \leq \! \int_{0}^{\tau}\! \Delta A_C \, \sqrt{\info^C} dt
\end{align} 
  bounds the maximum deviation that coherent dynamics can induce to    an incoherent process driving an observable $A$. The looser bound $\big| \fancya(\tau) \nobreak-\nobreak \fancya_I(\tau) \big| \nobreak \leq \nobreak 2 \sqrt{ \mathcal{J}_C(\rho_i,\rho_f) \frac{1}{\tau}  \int_{0}^{\tau}  \left( \Delta A_C\right)^2 } dt$  
   holds in terms of the quantum 
     divergence $\mathcal{J}_C(\rho_i,\rho_f) \nobreak \coloneqq \nobreak \tau \int_0^\tau \info^C dt$ 
     of the path.

   The maximum coherent speedup at any given time occurs when the coherent speed limit is saturated, e.g., when $\slogq \propto A_C$ (see Section~\ref{sec:saturation}). Using Eq.~\eqref{eq:saturationV2quant}, we prove in Appendix~\ref{app:qspeedup} that the Hamiltonian
   \begin{align}
\label{eq:Qspeed}
H_t^{\qs} \coloneqq - \lambda_t \frac{i}{2}  \sum_{j \neq k}^d  \frac{(p_j + p_k)}{p_j - p_k} A_{jk} \ket{j}\!\bra{k}
\end{align}
drives observable $A$ at such a speed limit. Here, the prefactor $\lambda_t$ sets the energy scale and would typically be determined by the available resources.
 The Hamiltonian $H_t^\qs = H_t^\qs\left(\rho_t,A\right)$ is tailored to the state of the system and the observable of interest, and makes an optimal use of the energetic resources to coherently drive the observable at its speed limit and, as a result, change it by a value $\big| \fancya_C(\tau) \big| \nobreak = \nobreak \int_{0}^{\tau}\! \Delta A_C \, \sqrt{\info^C} dt \nobreak = \nobreak \int_{0}^{\tau} \frac{\info^C}{\lambda_t}  dt$ within a time $\tau$.

In Appendix~\ref{app:qspeedup}, we illustrate how to exploit $H_t^\qs$ to enhance an incoherent process that erases information stored in a qubit.
We consider a system initialized in state $\ket{\Psi_0} \nobreak=\nobreak a\ket{0} \nobreak + \nobreak b \ket{1}$, with $a$ and $b$ real for simplicity, and assume that it is critical to hide from an adversary the fact that $z_0 \nobreak \coloneqq \nobreak \langle \sigma_z \rangle(0) \nobreak= a-b$ is not equal to zero. 
An incoherent dynamics, modelled by a Lindblad master equation $\dot\rho_t = \gamma \sum_{l = 0,1} \left( L_l \rho_t L_l^\dag - \tfrac{1}{2} \{ L_l^\dag L_l,\rho_t \}\right)$ with jump operators $L_0 = \ket{r}\!\bra{1}$ and $L_1 = \ket{r}\!\bra{0}$, drives the system to an orthogonal state $\ket{r}$ at a rate $\gamma$. 
Since $\bra{r}\! \sigma_z\! \ket{r} = 0$, this dynamics incoherently erases $z_0$ at a rate $\gamma$.
We find that the optimal Hamiltonian~\eqref{eq:Qspeed} that coherently enhances such a process 
is given by $H_t^{\qs} \nobreak = \nobreak \epsilon \, \tfrac{\sign(z_t)}{  \sign(x_t)} \, \sigma_y$, where $\epsilon \nobreak \coloneqq \nobreak \big\|H_t^\qs \big\|$ is set by the available energetic resources. 
While the purely incoherent dynamics hides that $z_0 \neq 0$ at a rate $-\dot z_t^{\textnormal{incoh}}/ z_t^{\textnormal{incoh}} \nobreak = \nobreak \gamma$, the unitarily-enhanced process leads to a faster rate of $-\dot z_t  / z_t \nobreak =\nobreak \gamma \nobreak+\nobreak 2 \epsilon |x_t/z_t|$.

 In this way, the new speed limits~\eqref{eq:bound2},~\eqref{eq:bound2v2}, and~\eqref{eq:boundlower}, and the bounds~\eqref{eq:totalchangeincoh} and~\eqref{eq:totalchangecoh},
pave the way to a systematic study of quantum speedups by identifying speed limits due to i) incoherent dynamics, ii) unitary quantum dynamics, and iii) arbitrary quantum dynamics, and in doing so, to a better understanding of the regimes in which enhancements due to coherent dynamics occur.

\section{Conclusions} \label{sec:discussion}
We derived speed limits on expectation values of 
observables for a quantum system evolving under arbitrary differentiable dynamics. These bounds  distinguish between classes of observables $A_Q$ driven solely by the Hamiltonian of a system and classes of  observables $A_C$ driven solely by incoherent dynamics. 
An observable will typically have coherent and incoherent contributions, $A = A_C + A_I$,
 and its speed will be bounded by a linear combination of the coherent Fisher information $\info^C$ and the incoherent Fisher information $\info^I$, weighted by associated fluctuations in the contributions to the observable.

The division of dynamics in terms of coherent and incoherent contributions in this way was crucial to deriving tighter bounds than those previously considered in the literature.
This division also allowed us to
i) prove lower bounds on the speed of evolution, 
ii) quantify 
the effects from non-unitary open dynamics of a quantum system, 
and 
iii) quantify the speedups that coherent dynamics can provide to incoherent processes. 

We expect these advances to broaden the field of applications of quantum speed limits 
by better capturing the timescales that are involved in the dynamics of different physical system observables.  
Our work also takes a step towards speed limits that correctly capture the dynamics of many-body systems, where it is known that speed limits are largely loose in estimating relevant time scales, e.g., for thermalization of quantum systems~\cite{Eisert_2015,Gogolin_2016}.
 A particularly interesting avenue for future research is that of incorporating constraints on the dynamics of physical systems, such as locality~\cite{liebrobinson1972}, integrability~\cite{rigol2008thermalization}, or those imposed by limited controllability~\cite{ControllabilityPRA2001,
 ControllabilityWerschnik_2007,
 SelsPRX2019,
 PoggiarXiv2020}.

\

\section*{Acknowledgements}
LPGP is thankful for extensive feedback from Jake Bringewatt, and for interesting discussions with Igor Boettcher, Lucas Brady, Stefano Cusumano, Yi-Kai Liu, {\L}ukasz Rudnicki, and Oles Shtanko.  
This work was supported by AFOSR MURI project ``Scalable Certification of Quantum Computing Devices and Networks'',
DoE FAR-QC project ``Fundamental Algorithmic Research for Quantum Computing'',
DoE ASCR Quantum Testbed Pathfinder program (award No.~DE-SC0019040), DoE ASCR Accelerated Research in Quantum Computing program (award No.~DE-SC0020312), U.S.~Department of Energy Award No.~DE-SC0019449,  NSF PFCQC program, AFOSR, ARO MURI, AFOSR MURI, PID2019-109007GA-I00,  the John Templeton Foundation and the Spanish  Ministerio  de Ciencia e Innovaci\'on (PID2019-109007GA-I00), and the National Science Foundation under
Grant No.~1856250.

\bibliography{referencesQSL}

\widetext
\clearpage

\appendix 


 

 
 \part*{ 
 \begin{center}
 \normalsize{
 APPENDIXES 
 } 
 \end{center}
 }
 
 
 In the Appendixes, we include detailed proofs of the results in the main text.
 
\begin{itemize}
\item Appendix~\ref{app:qPrice} --- we derive Eq.~(\ref{eq:generaldyn}) of Sec.~\ref{sec:generalQSL} in the main text, i.e., we show that the change in the expectation value of an observable is characterized by the covariance between the observable and the symmetric logarithmic derivative. 
    \vspace{-5pt}
 \item  Appendix~\ref{app:support} ---  we study the effect that a change in the support of the state has on the speed of an observable. 
We show that speed limit~\eqref{eq:generalbound} in the main text 
 is exact for states with constant support  and obtain an estimate of the error to the bound in cases when the rank of the state changes and the incoherent Fisher information diverges.
 \vspace{-5pt}
 \item Appendix~\ref{app:cohincoh} --- we prove speed limits on the coherent and incoherent contributions of an observable, Eqs.~\eqref{eq:bound2q},~\eqref{eq:bound2c}, and~\eqref{eq:bound2v2} of Sec.~\ref{sec:coherent-incoherent} in the main text.  
\vspace{-5pt}
\item  Appendix~\ref{app:FisherInformtions} --- 
we derive an upper bound on the incoherent Fisher information for a system interacting with an environment, Eq.~(\ref{eq:fisherbounds}b) in Sec.~\ref{sec:boundsfisher} of the main text.
\vspace{-5pt}
\item  Appendix~\ref{app:SIE} ---  focuses on speed limits to the von Neumann entropy, which results in a simple proof of the Small Incremental Entangling theorem in the ancilla-free case.
\vspace{-5pt}
\item  Appendix~\ref{app:comparison} ---  we compare the novel coherent-incoherent bound~\eqref{eq:bound2v2} with bound~\eqref{eq:generalbound}, derivable from the Cram\'er-Rao inequality, and show that the former is tighter than the latter by proving Eq.~\eqref{eq:saturation} of Sec.~\ref{sec:saturation} in the main text.
\vspace{-5pt}
\item  Appendix~\ref{app:saturation} ---  
we present detailed derivations for the last two paragraphs of Sec.~\ref{sec:saturation} and for Fig.~\ref{fig:fig2} in the main text. Specifically, we  construct operators that saturate the coherent and incoherent bounds for a qubit suffering dephasing and use these results to compare bounds~\eqref{eq:generalbound},~\eqref{eq:bound2}, and~\eqref{eq:bound2v2} in the main text.
\vspace{-5pt}
 \item Appendix~\ref{app:Divergence} ---  we prove Eq.~\eqref{eq:integrated} of Sec.~\ref{sec:geometry} in the main text, which gives a   bound on integrated changes in an observable in terms of the integrated Fisher information, also known as the \emph{divergence of a path} in state space.
\vspace{-5pt}
\item  Appendix~\ref{app:qspeedup} ---   we derive the Hamiltonian in Eq.~\eqref{eq:Qspeed} of the main text, which drives an observable at the maximum allowed speed. We show how it can be used to enhance a simple incoherent process on a qubit.
\vspace{-5pt}
\item Appendix~\ref{app:qslHilbert} --- includes a bound to the rate of change of the fidelity, as means to derive speed limits in state space from the coherent-incoherent speed limits. 
 \vspace{-5pt}
\item Appendix~\ref{app:ParamEst} ---  we discuss different connections between 
speed limits~\eqref{eq:bound2}, and~\eqref{eq:bound2v2} in Secs.~\ref{sec:generalQSL} and~\ref{sec:coherent-incoherent} of the main text, 
and bounds that are derivable from Cram\'er-Rao type bounds.
\end{itemize}

\section{Equation of motion for the expectation value of observables}
\label{app:qPrice}

In this Appendix, we derive Eq.~(\ref{eq:generaldyn}) of Sec.~\ref{sec:generalQSL} in the main text, i.e., we show that the change in the expectation value of an observable is characterized by the covariance between the observable and the symmetric logarithmic derivative.

\

Since the symmetric logarithmic derivative $\slog$ is implicitly defined by the expression $ \tfrac{d \rho_t}{dt} \eqqcolon \tfrac{1}{2}\{ \rho_t,\slog \} $ where $\{A,B \} = AB+BA$ denotes an anticommutator, we see that the change in an observable due to changes in the state satisfies
 \begin{align}
\label{eq-app:generaldynAux}
  \tr{\frac{d\rho_t}{dt} A} &= \tr{\frac{1}{2}\{ \rho_t,\slog \} A}  = \frac{1}{2} \tr{\rho_t \{ \slog,A\}}  = \frac{1}{2} \tr{\rho_t \{ \slog,A\}}  - \langle \slog \rangle \langle A \rangle \nonumber \\
&= \cov\left( \slog,A \right)   \eqqcolon  \dota,  
\end{align}
where we use the facts that the trace is cyclic and that $\langle \slog \rangle = \tr{\rho_t \slog} = \tfrac{1}{2} \tr{\{\rho_t, \slog\}} = \tr{\tfrac{d\rho_t}{dt}} = 0$ for trace preserving dynamics. Here, $\cov(A,\slog) \nobreak \coloneqq \nobreak \frac{1}{2}  \tr{\rho_t \{A,\slog\}} - \langle A \rangle \langle \slog \rangle$ is the symmetrized covariance between operators $\slog$ and $A$. 

The expectation value $\langle A \rangle = \tr{A\rho_t}$  
thus follows the equation of motion
\begin{align}
\label{eq-app:generaldyn}
\frac{d\langle A \rangle}{dt} &= \tr{\frac{d\rho_t}{dt} A} + \left\langle \frac{dA}{dt} \right\rangle   = \cov\left( \slog,A \right) + \left\langle \frac{dA}{dt} \right\rangle \eqqcolon  \dota + \left\langle \frac{dA}{dt} \right\rangle,
\end{align}
which is  Eq.~(\ref{eq:generaldyn}) of the main text.
 
\section{Assumptions on the support of the state}
\label{app:support}

In this Appendix, we study the effect that a change in the rank of the state has on the speed of an observable. Specifically, we show that bound~\eqref{eq:generalbound} in the main text is exact for states with constant support and obtain an estimate of the error to the bound in cases when the rank of the state changes and the incoherent Fisher information diverges.

 \
 
In deriving the first general bound [Eq.~(\ref{eq:generalbound}) in the main text] 
\begin{align}
\label{eq-app:generalbound}
\big| \dota \big| &\leq \Delta A \sqrt{\info},
\end{align}
we 
disregarded the contribution of levels $j$ such that $p_j = 0$.
 Here, we quantify the error introduced by this and argue that such an assumption gives a good approximation for most times for differentiable continuous evolution. 

Defining $\delta A \coloneqq A - \langle A \rangle$, we find
\begin{align}
\big| \dota \big| = \left| \tr{\delta A \frac{d\rho_t}{dt} } \right| = \left| \sum_{jk} \delta A_{jk} \bra{k} \frac{d\rho_t}{dt} \ket{j} \right| &=  \left| \sum_{p_j, p_k \neq 0} \delta A_{jk} \bra{k} \frac{d\rho_t}{dt} \ket{j} + \sum_{p_j, p_k = 0} \delta A_{jk} \bra{k} \frac{d\rho_t}{dt} \ket{j} \right| \nonumber \\
&\leq  \left| \sum_{p_j, p_k\neq 0} \delta A_{jk} \bra{k} \frac{d\rho_t}{dt} \ket{j} \right| + \left| \sum_{p_j, p_k = 0} \delta A_{jk} \bra{k} \frac{d\rho_t}{dt} \ket{j} \right|,
\end{align}
where we set aside the sum of terms with $p_j = 0$ and denoted matrix elements by $\bra{k} \delta A \ket{j} = \delta A_{jk}$. 
The first term includes all contributions to the rate of change that satisfy $p_j \neq 0$, as assumed in the main text, giving rise to Eq.~\eqref{eq-app:generalbound}, which is also Eq.~(\ref{eq:generalbound}) in the main text. 
The second term thus contains all factors that were disregarded in the bound $\big| \dota \big| \leq \Delta A \sqrt{\info}$ when states with $p_j = 0$ exist.

For the second term, using the fact that any trace-preserving differentiable evolution can be expressed as
\begin{align}
\dt \rho_t = - i[H_t,\rho_t] + U_t \frac{d \chi_t}{dt} U_t^\dag,
\end{align}
and that $\ket{j}\nobreak= \nobreak U_t  \ket{j}_0$ and $\chi_t = \sum_j p_j \ket{j}_0 \prescript{}{0\!}{\bra{j}}$, we find that
\begin{align}
\sum_{p_j, p_k = 0} \delta A_{jk} \bra{k} \tfrac{d\rho_t}{dt} \ket{j} &= \sum_{p_j, p_k = 0} \delta A_{jk} \bra{k} U_t \frac{d \chi_t}{dt} U_t^\dag \ket{j} = \sum_{p_j, p_k = 0} \delta A_{jk} \bra{k} U_t \sum_l \dot p_l  \ket{l}_0 \prescript{}{0\!}{\bra{l}} U_t^\dag \ket{j} \nonumber \\
&= \sum_{p_j, p_k = 0} \sum_l \dot p_l \delta A_{jk} \bra{k} l \rangle   \langle l \ket{j} = \sum_{p_j = 0} \dot p_j \, \delta A_{jj} \nonumber \\
&= \sum_{\{p_j = 0\}} \dot p_j A_{jj}.
\end{align}
This term, which would result in a  correction to 
Eq.~\eqref{eq:bound2c} in the main text, 
contributes very little
to the rate of change of $\dota$ for continuous evolutions 
because it is nonzero only 
for infinitesimal moments in time in which a non-populated state $j$ acquires a non-zero probability $p_j$. 
Whenever a state has constant support, the error term is zero.

\section{Coherent and incoherent contributions to speed limits for observables}
\label{app:cohincoh}

In this Appendix, we derive speed limits on the coherent and incoherent contributions of an observable, proving Eqs.~\eqref{eq:bound2q},~\eqref{eq:bound2c}, and~\eqref{eq:bound2v2} of Sec.~\ref{sec:coherent-incoherent} in the main text. 

\

The dynamics of any quantum system can be decomposed in terms of coherent and incoherent contributions as
\begin{align}
\label{eq-app:statedynamics}
\dt \rho_t = - i[H_t,\rho_t] + U_t \frac{d \chi_t}{dt} U_t^\dag,
\end{align}
where the Hamiltonian $H_t \coloneqq i \tfrac{d U_t}{dt} U_t^\dag$ drives the coherent dynamics and where $\chi_t =  \sum_j p_j(t)  \ket{j}_0 \prescript{}{0\!}{\bra{j}}$, the eigenvalues of $\rho_t \nobreak = \nobreak \sum_j p_j \ket{j}\bra{j}$ are $\{ p_j(t) \}$, and $\ket{j}\nobreak=\nobreak U_t  \ket{j}_0$.

Similarly, different components of an observable will evolve under the coherent and incoherent contributions to the dynamics. 
We define a coherent-incoherent separation of the observable of interest $A\nobreak\coloneqq\nobreak A_C\nobreak+\nobreak A_I$, with
\begin{align}
A_C \coloneqq \sum_{j\neq k} A_{jk} \ket{j}\!\bra{k} \qquad \& \qquad A_I \coloneqq \sum_{j} A_{jj} \ket{j}\!\bra{j}.
\end{align} 

The change $\dota_C \coloneqq \tr{\frac{d \rho_t}{dt} A_C}$
of the observable's coherent component is
\begin{align}
\label{eq-app:proofbound}
 \dota_C &=  \sum_{j \neq k}  A_{jk} \bra{k} \tfrac{d\rho_t}{dt} \ket{j}   =  -i \sum_{j \neq k}     A_{jk} \bra{k} [H_t,\rho_t] \ket{j}    +   \sum_{j \neq k}   A_{jk} \bra{k} U_t \frac{d \chi_t}{dt} U_t^\dag \ket{j}    \nonumber \\
&=  -i \sum_{j \neq k}     A_{jk} \frac{ \bra{k} [H_t,\rho_t] \ket{j} }{p_j + p_k} (p_j + p_k)     +   \sum_{j \neq k}  A_{jk}  \prescript{}{0\!}{\bra{k}} \Big( \sum_l \dot p_l(t)  \ket{l}_0 \prescript{}{0\!}{\bra{l}} \Big)  \ket{j}_0  \nonumber \\
&= \frac{1}{2}\tr{A_C \slogq \rho_t} + \frac{1}{2}\tr{A_C \rho_t \slogq }  +   \sum_{j\neq k}   A_{jk} \, \dot p_j(t) \delta_{jk}  \nonumber \\
&=\cov\left(A_C,\slogq \right)  ,
\end{align}
where we used the fact that the diagonal components $A_{jj}$ of the coherent component $A_C$ of the observable are zero, and that conservation of probability gives $\langle \slogq \rangle = 0$, where 
\begin{align}
\slogq \coloneqq -2i\sum_{j \neq k}    \frac{\bra{j}\nobreak [H_t,\rho_t] \ket{k}}{(p_j\nobreak+\nobreak p_k)} \ket{j} \! \bra{k} = -2i\sum_{j \neq k}    \frac{\bra{k}\nobreak [H_t,\rho_t] \ket{j}}{(p_j\nobreak+\nobreak p_k)} \ket{k} \! \bra{j}.
\end{align}

The speed limit follows from the Cauchy-Schwarz inequality:
\begin{align}
\label{eq-app:boundaux}
\left|  \dota_C  \right| &= \left|  \cov\left(A_C,\slogq \right)  \right| = \frac{1}{2} \left| \tr{\rho_t \delta A_C  \delta \slogq } + \tr{\rho_t \delta \slogq \delta A_C }  \right| \nonumber \\
&\leq \frac{1}{2} \Big( \sqrt{\tr{\rho_t (\delta A_C)^2} \tr{\rho_t (\delta \slogq)^2}} + \sqrt{ \tr{\rho_t (\delta A_C)^2} \tr{\rho_t (\delta \slogq)^2} } \Big) = \Delta A \, \Delta \slogq,
\end{align}
where we define $\delta A_C \coloneqq A_C - \langle A_C \rangle$ and $\delta \slogq \coloneqq \slogq - \langle \slogq \rangle$.

A direct calculation further shows that
\begin{align}
\Delta \slogq = \tr{\rho_t \slogq^2} = \info^C = 2 \sum_{j \neq k}  \frac{ \Big|\! \bra{j} [H_t,\rho_t] \ket{k} \!\Big|^2 }{p_j + p_k}
\end{align}
is the coherent contribution to the quantum Fisher information. 

This then
proves  Eq.~\eqref{eq:bound2q} in the main text:
 \begin{align}
 \label{eq-app:bound}
\big| \dota_C \big| &= \left|  \cov\left(A,\slogq \right)  \right| \leq \Delta A \, \Delta \slogq =\Delta A_C  \sqrt{ \info^C} .
\end{align}

For the evolution of the incoherent part, we use the fact that $A_I$ is diagonal in the basis of $\rho_t$, so that
\begin{align}
 \dota_I   &=   \sum_{jk} \delta A_{jj} \bra{k} \tfrac{d\rho_t}{dt} \ket{j}   =  \sum_{j}    \delta A_{jj} \bra{j} U_t \frac{d \chi_t}{dt} U_t^\dag \ket{j}     =    \sum_{j}    \delta A_{jj} \prescript{}{0\!}{\bra{j}} \Big( \sum_l \dot p_l(t)  \ket{l}_0 \prescript{}{0\!}{\bra{l}} \Big)  \ket{j}_0     \nonumber \\
&=   \sum_{j}    \delta A_{jj} \, \dot p_j(t)   .
\end{align}
This becomes identical to the expression for the change in a classical observable acting on a classical stochastic system.
 It was shown in~\cite{nicholson2020timeinformation} that dynamics of classical observables under stochastic dynamics satisfies a speed limit that depends on the classical Fisher information
$
\info^I \coloneqq \sum_j p_j \left( \frac{d}{dt} \ln p_j \right)^2. 
$

Defining
\begin{align}
\slogc &\coloneqq \sum_{j}  \frac{d \ln p_j}{dt} \ket{j} \! \bra{j} 
\end{align}
and disregarding states with $p_j = 0$ (see Appendix~\ref{app:support} for an expression of the error introduced by this), we get
\begin{align}
\dota_I = 
\sum_{j}    \dot p_j(t) \,  A_{jj} &= \sum_j p_j \frac{\dot p_j}{p_j}  A_{jj} = \sum_j p_j \dt \ln(p_j)   A_{jj} = \tr{\rho_t \slogc A} = \cov(A_I,\slogc), 
\end{align}
where we used the fact that $\langle \slogc \rangle = 0$.
The Cauchy-Schwarz inequality then gives 
\begin{align}
\big| \dota_I \big| = \left| \cov(A_I,\slogc) \right| \leq \Delta A_I \, \Delta \slogc = \Delta A_I \sqrt{\info^I}, 
\end{align}
which proves Eq.~\eqref{eq:bound2c} in the main text.

 Combining the results gives
 \begin{align}
\label{eq-app:bound2}
\big| \dota \big|  &\leq \big| \dota_C \big|+ \big| \dota_I \big| \leq  \Delta A_C \sqrt{\info^C} + \Delta A_I \sqrt{\info^I},
\end{align}
proving the coherent-incoherent decomposition of the central bound in the main tex, Eq.~\eqref{eq:bound2v2}.

The reverse triangle inequality~\cite{Abramowitz} implies that $\left| x + y \right| \geq \left| x \right| - \left| y \right|$. Combining this with the fact that $\dota = \dota_Q + \dota_C$ and with bounds~\eqref{eq-app:bound} and~\eqref{eq-app:bound2} gives
\begin{align}
\big| \dota \big|  &= \big| \dota_C + \dota_I \big| \geq \big| \dota_C \big| - \big| \dota_I \big| \geq  \big| \dota_C \big| - \Delta A_I \sqrt{\info^I}, \\
\big| \dota \big|  &= \big| \dota_C + \dota_I \big| \geq \big| \dota_I \big| - \big| \dota_C \big| \geq  \big| \dota_I \big| -\Delta A_C \sqrt{\info^C}.
\end{align}
This proves the lower speed limit in Eq.~(\ref{eq:boundlower}) of the main text.

Finally, we note that the quantum Fisher information defined by Eq.~\eqref{eq:qFisher} in the main text can be decomposed in terms of the coherent and incoherent contributions to the dynamics, $\info = \info^C + \info^I$, with
\begin{align}
\info^C \coloneqq \Delta \slogq = 2 \sum_{j \neq k}  \frac{ \Big|\! \bra{j} [H_t,\rho_t] \ket{k} \!\Big|^2 }{p_j + p_k}, \qquad
\info^I \coloneqq \Delta \slogc = \sum_j p_j \left( \frac{d}{dt} \ln p_j \right)^2.
\end{align}

\section{Upper bound on the incoherent Fisher information} 
\label{app:FisherInformtions}

In this Appendix, we derive an upper bound on the incoherent Fisher information for a system interacting with an environment. 
We prove the second part of Eq.~\eqref{eq:fisherbounds} in Sec.~\ref{sec:boundsfisher} of the main text and we prove that the bound is loose. 

\

The quantum Fisher information is $\info = \info^C + \info^I$, where
\begin{align}
\info^C \coloneqq  2 \sum_{j \neq k}  \frac{ \Big|\! \bra{j} [H_t,\rho_t] \ket{k} \!\Big|^2 }{p_j + p_k}, \qquad
\info^I \coloneqq \sum_j p_j \left( \frac{d}{dt} \ln p_j \right)^2.
\end{align}
Braunstein and Caves proved an upper bound to the  coherent Fisher information
\begin{align}
\info^C \leq 4 (\Delta H_t)^2,
\end{align}
where $H_t$ is the Hamiltonian of the system~\cite{BraunsteinCaves1994}. Equality holds only if the state is pure or in the trivial case $H_t \propto \id$.

Here, we prove an analogous bound for the incoherent Fisher information,
\begin{align}
\info^I = \sum_j p_j \left( \frac{d}{dt} \ln p_j \right)^2 = \sum_j \frac{(\dot p_j)^2}{p_j}.
\end{align}
Let us assume that the incoherent dynamics stems from the interaction between the system and a second system, i.e., an environment. The two interact via a Hamiltonian $H_t^\textnormal{int}$.
That is, if $H_t$ and $H_t^\env$ represent the Hamiltonians of the system of interest and the environment, the total system-environment Hamiltonian is $H_t^{\sys \env} = H_t \otimes \id_\env + \id_\sys \otimes  H_t^\env + H_t^\textnormal{int}$. Note that self Hamiltonians $H_t$ and $H_t^\env$
do not change the eigenvalues of the state of the system, so $H_t^\textnormal{int}$ is the only source of $\info^C \neq 0$.  We assume that $H_t^\textnormal{int}$ has support on the Hilbert space of the system and the environment, and acts on the state causing incoherent dynamics of the system. This excludes, for instance, cases in which the system or the
 environment are in eigenstates of $H_t^\textnormal{int}$.

Then, we have
\begin{align}
\dot p_j  &= \bra{j} \frac{d\rho_t}{dt} \ket{j} 
= \bra{j} -i [H_t,\rho_t] -i \trs{\env}{\left[ H_t^\textnormal{int},\rho_t^{\sys \env} \right]} \ket{j} \nonumber \\
&= -i  \sum_e \bra{j}\!  \bra{e}  \left[ H_t^\textnormal{int},\rho_t^{\sys \env} \right] \ket{e}\! \ket{j} ,
\end{align}
where $\rho_t^{\sys \env}$ is the joint state of the system $\sys$ and the environment $\env$, $\rho_t \coloneqq \trs{\env}{\rho_t^{\sys \env}} = \sum_e \bra{e} \rho_t^{\sys \env} \ket{e}$ is the state of the system, and $\{ \ket{e} \}$ denotes an arbitrary basis in the Hilbert space of $\env$. Defining the projector $X_j \equiv \sum_e \ket{e}\!\ket{j}\!\bra{j}\!\bra{e}$ and the shifted Hamiltonian $\delta H_t^\textnormal{int} \coloneqq H_t^\textnormal{int} - \tr{H_t^\textnormal{int} \rho_t^{\sys \env}}$, we get
\begin{align}
\label{eq-app:proofboundFisher}
\info^I &= \sum_j \frac{(\dot p_j)^2}{p_j} = \sum_j \frac{\left| \sum_e \bra{j}\!  \bra{e}  \left[ H_t^\textnormal{int},\rho_t^{\sys \env} \right] \ket{e}\! \ket{j} \right|^2}{p_j}  = \sum_j \frac{\left| \tr{X_j \left[\delta  H_t^\textnormal{int},\rho_t^{\sys \env} \right] } \right|^2}{p_j} \nonumber \\
&\leq \sum_j \frac{\left| 2 \tr{X_j  \delta  H_t^\textnormal{int} \rho_t^{\sys \env} } \right|^2}{p_j}  = 4\sum_j \frac{\left|   \tr{X_j  \delta  H_t^\textnormal{int} \sqrt{ \rho_t^{\sys \env} } \sqrt{ \rho_t^{\sys \env} } X_j} \right|^2}{p_j} \nonumber \\ 
&\leq 4\sum_j \frac{   \tr{X_j  \delta  H_t^\textnormal{int} \rho_t^{\sys \env} \delta  H_t^\textnormal{int} X_j } \tr{X_j \rho_t^{\sys \env} X_j }  }{p_j} 
= 4\sum_j \frac{   \tr{ X_j \delta  H_t^\textnormal{int} \rho_t^{\sys \env} \delta  H_t^\textnormal{int} } \tr{X_j \rho_t^{\sys \env} }  }{p_j} \nonumber \\
&= 4\sum_j \frac{   \tr{ \sum_e \ket{e}\!\ket{j}\!\bra{j}\!\bra{e} \delta  H_t^\textnormal{int} \rho_t^{\sys \env} \delta  H_t^\textnormal{int} } \tr{\sum_{e'} \ket{e'}\!\ket{j}\!\bra{j}\!\bra{e'} \rho_t^{\sys \env} }  }{p_j}  \nonumber \\
&= 4\sum_j \frac{   \bra{j }\trs{\env}{  \delta  H_t^\textnormal{int} \rho_t^{\sys \env} \delta  H_t^\textnormal{int} } \ket{j} \bra{j} \trs{\env}{  \rho_t^{\sys \env} } \ket{j} }{p_j} = 4\sum_j \frac{   \bra{j }\trs{\env}{  \delta  H_t^\textnormal{int} \rho_t^{\sys \env} \delta  H_t^\textnormal{int} } \ket{j} \bra{j} \rho_t \ket{j} }{p_j} \nonumber \\
&= 4\sum_j  \bra{j }\trs{\env}{  \delta  H_t^\textnormal{int} \rho_t^{\sys \env} \delta  H_t^\textnormal{int} } \ket{j} = 4 \tr{\left(\delta H_t^\textnormal{int} \right)^2 \rho_t^{\sys \env}} = 4 \left(\Delta H_t^\textnormal{int} \right)^2.
\end{align}
We used the fact that $|\tr{ABC}| = |\tr{ACB}|$ holds for Hermitian operators in the third line. We also used the facts that $X_j$ is a projector and $\rho_t^{\sys \env}$ is positive to apply the Cauchy-Schwarz inequality on line four. 
The variance of the interaction Hamiltonian between the system and the environment is $\left(\Delta H_t^\textnormal{int}\right)^2 \coloneqq \tr{\left(\delta H_t^\textnormal{int} \right)^2 \rho_t^{\sys \env}}$.
That is, we get the second part of Eq.~(\ref{eq:fisherbounds}) in the main text:
\begin{align}
\info^I \leq 4 \left(\Delta H_t^\textnormal{int} \right)^2,
\end{align}
mirroring the bound on the coherent Fisher information in terms of the variance of the system Hamiltonian, i.e., the first part of Eq.~(\ref{eq:fisherbounds}) in the main text.

The proof of Eq.~\eqref{eq-app:proofboundFisher} involves the Cauchy-Schwarz inequality between the operators $V_j \coloneqq X_j  \delta  H_t^\textnormal{int} \sqrt{ \rho_t^{\sys \env} }$ and $W_j \coloneqq X_j \sqrt{ \rho_t^{\sys \env} } $, which is tight if and only if $V_j = \frac{\tr{V_j W_j^\dag}}{\tr{W_j W_j^\dag}}W_j$ or in the trivial case when one of the operators is null. We now prove that the former is never the case. We have that
\begin{align}
 \sum_j \tr{V_j \sqrt{\rho_t^{\sys \env}}} &=  \sum_j  \tr{ X_j  \delta  H_t^\textnormal{int}   \rho_t^{\sys \env}  } = \tr{ \delta  H_t^\textnormal{int}   \rho_t^{\sys \env}  } = 0 \\
\sum_j \tr{W_j \sqrt{\rho_t^{\sys \env}}} &=  \sum_j   \tr{ X_j \rho_t^{\sys \env} }  = \tr{ \rho_t^{\sys \env} } = 1 ,
\end{align}
 where we used that $\delta H_t^\textnormal{int} \coloneqq H_t^\textnormal{int} - \tr{H_t^\textnormal{int} \rho_t^{\sys \env}}$. This proves that one cannot have $V_j \propto W_j$ and that the bound in Eq.~\eqref{eq-app:proofboundFisher} is loose except in the trivial case when both sides of the inequality are null, as stated after Eq.~\eqref{eq:fisherbounds} in Sec.~\ref{sec:boundsfisher} of the main text.

\section{A corollary:  Small Incremental Entangling theorem without ancillas}
\label{app:SIE}
 
 In this Appendix, we focus on speed limits for the von Neumann entropy, which result in a simple proof of the Small Incremental Entangling theorem in the ancilla-free case stated below Eq.\ (\ref{eq:boundentropy}) in the main text: $|\dot S | \lesssim \ln d \, \| H_t^\textnormal{int} \|$. 

\

The von Neumann entropy can be expressed as the expectation value of the `suprisal' operator $[-\ln \rho_t]$, $S \nobreak=\nobreak \left\langle -\ln \rho_t \right\rangle = - \tr{\rho_t \ln \rho_t}$. 
Following~\cite{ReebIEEE2015,
HenrikarXiv2020}, the variance of the suprisal operator satisfies
\begin{align}
\left( \Delta S \right)^2 &\coloneqq \tr{\rho_t (\ln \rho_t )^2} - S^2  = \left( \ln(2) \right)^2 \left[  \tr{\rho_t (\log \rho_t )^2} - (\tr{\rho_t \log \rho_t})^2 \right] \nonumber \\
&\leq \left( \ln(2) \right)^2 \left[ \frac{(\log(d-1))^2}{4 } + \frac{1}{(\ln(2))^2} \right]  =   \frac{(\ln(d-1))^2}{4 } + 1,
\end{align} 
where we used the fact that $\ln(x) = \ln(2)\log(x)$, with $\log(x) \coloneqq \log_2(x)$. 
Combined with Eqs.~\eqref{eq:fisherbounds} and~\eqref{eq:boundentropy}, this gives  
\begin{align}
\big| \dot S \big| \leq \Delta S \, \sqrt{\info^I} \leq \sqrt{(\ln(d-1))^2 + 4} \, \Delta H_t^\textnormal{int},
\end{align}
where $d$ is the dimension of the Hilbert space of the system. In the limit $d \gg 1$, this is
\begin{align}
\big| \dot S \big| &\lesssim  \, \ln(d) \, \Delta H_t^\textnormal{int}  \leq \, \ln(d) \left\| H_t^\textnormal{int} \right\|,
\end{align}
where the spectral norm $\left\| H_t^\textnormal{int} \right\|$ is given by the largest eigenvalue of $H_t^\textnormal{int}$.
The last inequality recovers the scaling of the Small Entangling theorem in the ancilla-free case~\cite{BravyiPRA2007,
MarienCommMathPhys2016}.

\section{Comparison of the coherent-incoherent bound~\eqref{eq:bound2v2} and  bound~\eqref{eq:generalbound}} 
\label{app:comparison}

Here we compare the novel bound~\eqref{eq:bound2v2} with bound~\eqref{eq:generalbound}, derivable from the Cram\'er-Rao inequality. We show that the former is tighter than the latter, and 
prove Eq.~\eqref{eq:saturation} of Sec.~\ref{sec:saturation} in the main text.

\

Equation~\eqref{eq:generalbound} in the main text reads
\begin{align}
\label{eq-app:generalboundrepeat}
\big| \dota \big| \leq \Delta A \sqrt{\info}.
\end{align}
We will now prove that this speed limit is looser than the coherent-incoherent upper bound [Eq.~\eqref{eq:bound2v2} in the main text] which reads
\begin{align}
\label{eq-app:bound2repeat}
\big| \dota \big|  &\leq \big| \dota_C \big| + \big| \dota_I \big| \leq  \Delta A_C \sqrt{\info^C} + \Delta A_I \sqrt{\info^I},
\end{align}
where $\info = \info^C + \info^I$. 
To do this, we first note that $A = A_C + A_I$, with 
\begin{align}
A_C \coloneqq \sum_{j\neq k} A_{jk} \ket{j}\!\bra{k} \qquad \& \qquad A_I \coloneqq \sum_{j} A_{jj} \ket{j}\!\bra{j},
\end{align} 
which implies that $\tr{\rho_t A} = \tr{\rho_t A_I}$ and $\tr{\rho_t A_C} = 0$. Then
\begin{align}
(\Delta A)^2 &=   \tr{\rho_t (A_C + A_I)^2} - \left[ \tr{\rho_t(A_C + A_I)} \right]^2   \nonumber \\
&=   \tr{\rho_t  A_C^2} + \tr{\rho_t A_I^2} + 2\tr{\rho_t A_C  A_I } -  \left[ \tr{\rho_t  A_I } \right]^2   \nonumber \\
&= (\Delta A_C)^2 + (\Delta A_I)^2,
\end{align}
where in the last line we used the fact that $\tr{\rho_t A_C  A_I } = \sum_{jk}p_j (A_C)_{jk} (A_I)_{kj} = 0$, since $A_C$ has zero diagonal elements while $A_I$ is  non-zero only on the diagonal.  

Dividing the right-hand side of Eq.~\eqref{eq-app:generalboundrepeat} by the right-hand side of Eq.~\eqref{eq-app:bound2repeat} gives
\begin{align}
\frac{ \Delta A \sqrt{\info}  }{ \Delta A_C \sqrt{\info^C} + \Delta A_I \sqrt{\info^I} } & = \frac{ \sqrt{\left[ (\Delta A_C)^2 + (\Delta A_I)^2 \right] \left(\info^C + \info^I \right)}  }{ \Delta A_C \sqrt{\info^C} + \Delta A_I \sqrt{\info^I} }  = \sqrt{ \frac{ \left[ (\Delta A_C)^2 + (\Delta A_I)^2 \right] \left(\info^C + \info^I \right) }{ \left( \Delta A_C \sqrt{\info^C} + \Delta A_I \sqrt{\info^I} \right)^2 } } \nonumber \\
&= \sqrt{ \frac{ (\Delta A_C)^2 \info^C + (\Delta A_I)^2 \info^I + (\Delta A_C)^2 \info^I + (\Delta A_I)^2 \info^C }{  \left( \Delta A_C \sqrt{\info^C} + \Delta A_I \sqrt{\info^I} \right)^2  } } \nonumber \\
&= \sqrt{ \frac{ \left( \Delta A_C \sqrt{\info^C} + \Delta A_I \sqrt{\info^I} \right)^2 -2 \Delta A_C \Delta A_I \sqrt{\info^C \info^I} + (\Delta A_C)^2 \info^I + (\Delta A_I)^2 \info^C  }{ \left( \Delta A_C \sqrt{\info^C} + \Delta A_I \sqrt{\info^I} \right)^2 } } \nonumber \\
&=\sqrt{1+\frac{\left( \Delta A_C \sqrt{\info^I} - \Delta A_I \sqrt{\info^C} \right)^2}{\left( \Delta A_C \sqrt{\info^C} + \Delta A_I \sqrt{\info^I} \right)^2}} \geq 1,
\end{align}
proving Eq.~\eqref{eq:saturation} in the main text. This shows that
that the coherent-incoherent bound~\eqref{eq-app:bound2repeat} is tighter than~\eqref{eq-app:generalboundrepeat} [bound~\eqref{eq:bound2v2} is tighter than~\eqref{eq:generalbound} in the main text].
The bounds coincide only when $\Delta A_C \sqrt{\info^I}\nobreak =\nobreak \Delta A_I \sqrt{\info^C}$.

\section{Comparing the bounds on a qubit}
\label{app:saturation}

 In this Appendix, we present detailed derivations for the last two paragraphs of Sec.~\ref{sec:saturation} and for Fig.~\ref{fig:fig2} in the main text. Specifically, we  construct operators that saturate the coherent and incoherent bounds 
 for a qubit suffering dephasing and use these results to compare bounds~\eqref{eq:generalbound},~\eqref{eq:bound2}, and~\eqref{eq:bound2v2} in the main text.
 
\


Let us consider a qubit with Hamiltonian 
\begin{align}
H\nobreak = \nobreak \tfrac{\omega}{2} \sigma_y \nobreak 
\end{align}
and state 
\begin{align}
\rho_t = \frac{\id + x \sigma_x + z \sigma_z}{2},
\end{align}
constrained to the $y = \tr{\rho \sigma_y } = 0$ plane, 
with $x = \tr{\rho \sigma_x }$ and 
$z = \tr{\rho \sigma_z }$.

We have
\begin{align}
[H,\rho_t] = \frac{\omega}{2} \left( \tfrac{x}{2} [\sigma_y,\sigma_x] + \tfrac{z}{2} [\sigma_y,\sigma_z] \right) = i\frac{\omega}{2} \left( -x \sigma_z + z \sigma_x \right),
\end{align}
which means that 
\begin{align}
\slogq &= -2 i \sum_{j \neq k}    \frac{\bra{j}\nobreak [H_t,\rho_t] \ket{k}}{(p_j\nobreak+\nobreak p_k)} \ket{j} \! \bra{k}   =  \omega \sum_{j \neq k}    \frac{\bra{j}\nobreak \left( -x \sigma_z + z \sigma_x \right) \ket{k}}{(p_j\nobreak+\nobreak p_k)} \ket{j} \! \bra{k} = \omega \sum_{j \neq k}    \bra{j}\nobreak \left( -x \sigma_z + z \sigma_x \right) \ket{k} \ket{j} \! \bra{k} \nonumber \\
&= \omega \sum_{j  k}    \bra{j}\nobreak \left( -x \sigma_z + z \sigma_x \right) \ket{k} \ket{j} \! \bra{k} - \omega \sum_{j}    \bra{j}\nobreak \left( -x \sigma_z + z \sigma_x \right) \ket{j} \ket{j} \! \bra{j} \nonumber \\
&= \omega \nobreak \left( -x \sigma_z + z \sigma_x \right) - \omega \sum_{j}    \bra{j}\nobreak \left( -x \sigma_z + z \sigma_x \right) \ket{j} \ket{j} \! \bra{j}  ,
\end{align}
where we used the fact that, for a two-level system, $p_1+p_2 = 1$.
 
 Then, 
 the operator $\sigma_y$ satisfies
 \begin{align}
\cov\left( \slogq, \sigma_y \right) &= \frac{1}{2} \omega \tr{\rho_t \{  -x \sigma_z + z \sigma_x  , \sigma_y\}} -  \frac{1}{2} \omega \sum_{j }    \bra{j}\nobreak \left( -x \sigma_z + z \sigma_x \right) \ket{j} \tr{\rho_t \{ \ket{j}\!\bra{j}, \sigma_y \}}    = 0,   
\end{align}
where we used the fact that $\{\sigma_x,\sigma_y \} = \{\sigma_z,\sigma_y \} = 0$ and that
the eigenvectors $\{\ket{j}\}$ of $\rho_t$ belong to the $y = 0$ plane, which implies that $\tr{\rho_t \{ \ket{j}\!\bra{j}, \sigma_y \}}   = 0$. Note that this can also be concluded directly from the fact that under unitary dynamics  $\dot y = \cov(\slogq,\sigma_y)$ and that, for the chosen initial state and Hamiltonian, we have $\dot y = 0$.

Consider now a non-unitary term in the dynamics causing dephasing along $\sigma_z$ with a rate $\kappa$, modeled by $U_t \tfrac{d \chi_t}{dt} U_t^\dag \nobreak = \nobreak - \kappa [\sigma_z,[\sigma_z,\rho_t]]$ in Eq.~\eqref{eq:dynamics} in the main text. This leaves the qubit to be constrained to the $y = 0$ plane as well. 
The operator $\sigma_y$ then satisfies
\begin{align}
\cov(\slogc,\sigma_y) =  \sum_j p_j \frac{\dot p_j}{p_j}\bra{j} \sigma_y \ket{j} = \sum_j \dot p_j \bra{j} \sigma_y \ket{j} =0.
\end{align}

Given that $\cov(\slogq,\id) = \cov(\slogc, \id) = 0$, this implies that the set $\{ \slogq,\slogc,\sigma_y,\id \}$ forms a complete basis of orthogonal operators. The operators $\id$ and $\sigma_y$ are thus `still', i.e., they evolve neither under the Hamiltonian nor under the dephasing.
Moreover, since $\sigma_x$ and $\sigma_z$ are orthogonal to $\sigma_y$ and $\id$, the former two can be written solely
in terms of the preferred `speed' operators as
\begin{align}
\sigma_z = \alpha_C^z \slogq + \alpha_I^z \slogc, \qquad  \sigma_x = \alpha_C^x \slogq + \alpha_I^x \slogc.
\end{align}

Observables $\sigma_z$ and $\sigma_x$ thus saturate the coherent and incoherent bounds~\eqref{eq:bound2} in the main text. In the case of unitary dynamics ($\kappa = 0$), we have $\slogc = 0$ and $\slog = \slogq$, which means that the upper bound~\eqref{eq:generalbound} in the main text is also saturated. Finally, saturation of the upper bound~\eqref{eq:bound2v2} depends on the relative signs of $\alpha_C$ and $\alpha_I$: when $\sign(\alpha_C) \neq \sign(\alpha_I)$, Eq.~\eqref{eq:bound2v2} is not saturated.

\section{Bound in terms of path-divergences}
\label{app:Divergence}

In this Appendix, we derive bounds on integrated changes in an observable in terms of \emph{path-divergences}.
We prove Eq.~\eqref{eq:integrated} of Sec.~\ref{sec:geometry} in the main text and show how to upper bound Eqs.~\eqref{eq:totalchangeincoh} and~\eqref{eq:totalchangecoh} in terms of coherent and incoherent path-divergences.

\

The \emph{divergence of a path} is defined 
by~\cite{CrooksPRL2007,NicholsonPRE2018,GirolamiPRL2019}
\begin{align}
\mathcal{J}(\rho_i,\rho_f) \coloneqq \tau \int_0^\tau \info dt
\end{align}
and is related to the \emph{energy of a path} $\mathcal{J}/2\tau$.
Note that the square of the length of a path is upper bounded by the path's divergence: 
\begin{align}
  \mathcal{L}(\rho_i,\rho_f) &=   \int_{0}^{\tau} \sqrt{\info} dt =   \tau \frac{1}{\tau} \int_{0}^{\tau} \sqrt{\info} dt \leq  \tau \sqrt{\frac{1}{\tau}\int_{0}^{\tau} \info dt } =  \tau \sqrt{\frac{1}{\tau^2} \mathcal{J}(\rho_i,\rho_f)  } = \sqrt{\mathcal{J}(\rho_i,\rho_f)}.
\end{align}

Equations~\eqref{eq:generalbound},~\eqref{eq:buresmetric}, and the Cauchy-Schwarz inequality lead to an integrated bound
\begin{align}
\big| \fancya(\tau)   \big| &= \left| \int_{0}^{\tau} \dota \, dt \right| \leq 2 \int_{0}^{\tau} \sqrt{\info} \, \Delta A \, dt  = 2\tau \frac{1}{\tau} \int_{0}^{\tau} \sqrt{\info} \, \Delta A \, dt   \nonumber \\
&\leq 2\tau \sqrt{ \frac{1}{\tau} \int_{0}^{\tau}  \info  \, dt } \sqrt{ \frac{1}{\tau} \int_{0}^{\tau}  \left( \Delta A\right)^2 \, dt }.
\end{align}
Then, the total change in an observable is bounded by the divergence of the path in Hilbert space and the integrated fluctuations in the observable:
\begin{align}
\big| \fancya(\tau)   \big|^2 &\leq 4\tau^2 \left( \frac{1}{\tau} \int_{0}^{\tau}  \info  \, dt \right) \left( \frac{1}{\tau} \int_{0}^{\tau}  \left( \Delta A\right)^2 \, dt' \right) = 4 \mathcal{J}(\rho_i,\rho_f) \frac{1}{\tau}      \int_{0}^{\tau}  \left( \Delta A\right)^2 \, dt.
\end{align}
This proves Eq.~\eqref{eq:integrated} in the main text.

The divergence of a path can be decomposed into coherent and incoherent terms, $\mathcal{J}(\rho_i,\rho_f) = \mathcal{J}_C(\rho_i,\rho_f) + \mathcal{J}_I(\rho_i,\rho_f)$, with
\begin{align}
\mathcal{J}_C(\rho_i,\rho_f) \coloneqq \tau \int_0^\tau \info^C dt, \quad \textnormal{and} \quad  \mathcal{J}_I(\rho_i,\rho_f) \coloneqq \tau \int_0^\tau \info^I dt.
\end{align}
The total changes in the observable due to coherent and incoherent dynamics are bounded by 
\begin{align}
\big| \fancya(\tau) \nobreak-\nobreak \fancya_I(\tau) \big| \nobreak &\leq \nobreak 2 \sqrt{ \mathcal{J}_C(\rho_i,\rho_f) \frac{1}{\tau}  \int_{0}^{\tau}  \left( \Delta A_C\right)^2 } dt,   \\
\big| \fancya(\tau) \nobreak-\nobreak \fancya_C(\tau) \big| \nobreak &\leq \nobreak 2 \sqrt{ \mathcal{J}_I(\rho_i,\rho_f) \frac{1}{\tau}  \int_{0}^{\tau}  \left( \Delta A_I\right)^2 } dt.
\end{align}
This shows how to upper bound Eqs.~\eqref{eq:totalchangeincoh} and~\eqref{eq:totalchangecoh} in the main text in terms of coherent and incoherent path-divergences.

\section{Coherent speedup of incoherent processes }
\label{app:qspeedup}

In this Appendix, we derive the Hamiltonian in Eq.~\eqref{eq:Qspeed} of Sec.~\ref{sec:geometry} in the main text, which drives an observable at the maximum allowed speed. We then use this Hamiltonian to enhance an incoherent process on a qubit, 
presenting detailed derivations for the two paragraphs following Eq.~\eqref{eq:Qspeed}.

\

What is the Hamiltonian that induces the fastest change to an observable $A$? The speed with which the observable changes due to coherent drive satisfies $\dota_C = \cov(\slogq,A_C) \leq \Delta \slogq \, \Delta A_C$. Thus, any Hamiltonian that leads to a symmetric logarithmic derivative for which $\slogq \propto A_C$ will saturate the speed limit and drive the observable as fast as allowed by nature.  Using Eq.~\eqref{eq:saturationV2quant}, a direct calculation shows that the Hamiltonian [Eq.~\eqref{eq:Qspeed} in the main text]
\begin{align}
\label{eq:app-Hspeedup}
H_t^{\qs} \coloneqq - \lambda_t \frac{i}{2}  \sum_{j \neq k}  \frac{(p_j + p_k)}{p_j - p_k} A_{jk} \ket{j}\!\bra{k}
\end{align}
leads to 
\begin{align}
L_C^\qs \nobreak = \nobreak \lambda_t A_C.
\end{align}
This Hamiltonian thus induces dynamics for which the coherent bound for observable $A$ saturates. The prefactor $\lambda_t$ sets the energy scale of the Hamiltonian and would typically be constrained by the available resources.  
Note that the Hamiltonian $H_t^{\qs}$ that leads the observable to evolve at the speed limit is state- and observable-dependent---it is a Hamiltonian tailored to drive the expectation value of the observable as rapidly as possible given the available resources.

The observable evolves according to
\begin{align}
\dota_C = \cov(A_C,L_C) = \Delta A_C \, \Delta L_C^\qs  = \lambda_t \, \left(\Delta A_C\right)^2 =  \frac{ \info^C}{\lambda_t},
\end{align}
and the integrated change in the observable due to coherent dynamics becomes
\begin{align}
\label{eq-app:cohchange}
\big| \fancya(\tau) - \fancya_I(\tau) \big| = \left| \fancya_C(\tau) \right|  &= \left| \int_0^\tau \cov(A_C,\slogq^\qs) dt \right| = \int_0^\tau \lambda_t \, \left(\Delta A_C\right)^2 dt = \int_{0}^{\tau} \frac{\info^C}{\lambda_t}  dt.
\end{align} 
 This coherent drive can be used to enhance incoherent processes, up to an amount characterized by the integrated coherent Fisher information weighed by the scaling prefactor $\lambda_t$.
 
 We consider a toy model of an incoherent erasure process of a qubit~\cite{wilde2013quantum}.  In an auxiliary reset state $\ket{r}$, the expectation value of all qubit observables is null, e.g., $\langle \sigma_x \rangle \nobreak = \nobreak \langle \sigma_y \rangle \nobreak = \nobreak \langle \sigma_z \rangle  \nobreak = \nobreak 0 $. 
Incoherent dynamics drives the system to such an erased state 
with a rate $\gamma$. 
We model this incoherent dynamics by a Lindblad master equation with jump operators $L_0 \coloneqq \ket{r}\! \bra{0}$ and $L_1 \coloneqq \ket{r}\! \bra{1}$, 
\begin{align}
 \label{eq-app:erasingchannel}
 \dot \rho_t &= \gamma \sum_{j = 0,1} \left( L_j \rho_t L_j^\dag - \frac{1}{2} \{L_j^\dag L_j, \rho_t \}  \right) = \gamma \left( \tr{\rho_t \Pi_{2}}   \ket{r}\!\bra{r}  - \frac{1}{2} \{ \Pi_2 , \rho_t  \}  \right),
 \end{align}
where we define the projector $\Pi_2 \coloneqq \ket{0}\!\bra{0} + \ket{1}\!\bra{1}$ onto the subspace of the two-level system.

For illustration purposes, we suppose that it is more critical to erase certain aspects of the information stored in the initial state $\ket{\Psi_0} = a \ket{0} + b\ket{1}$ of the two-level system. For instance, this could be because it is expected that an adversary will attempt to acquire information about the initial value $z_0 \coloneqq \langle \sigma_z \rangle(0) = a-b$, by measuring $\sigma_z$ on an ensemble of such effective qubits, each one initially prepared in $\ket{\Psi_0}$. We wish to hide the fact that $z_0 \neq 0$. 

Under the incoherent erasure process, information of $z_0$ is exponentially erased at a rate $\gamma$, 
  \begin{align}
 \label{eq-app:incoherasure}
     \dot z_t^{\textnormal{incoh}} = \tr{\sigma_z \dot \rho_t} = \gamma \left( \tr{\rho_t \Pi_2} \bra{r}\! \sigma_z \!\ket{r} - \frac{1}{2} \tr{\{\sigma_z,\Pi_2 \} \rho_t \} } \right) = - \gamma \tr{\sigma_z \rho_t} =- \gamma \, z_t^{\textnormal{incoh}},
 \end{align}
 where we used that $\bra{r} \!\sigma_z\! \ket{r} = 0$.

We can take advantage of a coherent drive to enhance the process of hiding the fact that $z_0 \neq 0$ from the adversary. 
Hamiltonian $H_t^{\qs}$ in Eq.~\eqref{eq:app-Hspeedup} [Eq.~\eqref{eq:Qspeed} in the main text] defines the fastest time-local way to do this. Given that $\sigma_z\!\ket{r} = 0$, at time $t=0$ the Hamiltonian $H_t^{\qs}$ in Eq.~\eqref{eq:app-Hspeedup} does not connect the reset state $\ket{r}$ to states $\ket{0}$ and $\ket{1}$. This implies that $\ket{r}$ remains an eigenstate of $\rho_t$ under the action of the incoherent dynamics as well as the coherent dynamics. Moreover, since dynamics is coherent within the qubit subspace, we can denote the remaining eigenstates of  the evolved state by $\ket{\Psi_t}$ and $\ket{\Psi_t^\perp}$ with eigenvalues $p_\Psi$ and $0$. The state then takes the form $\rho_t \nobreak=\nobreak p_\Psi(t) \ket{\Psi_t}\!\bra{\Psi_t} \nobreak+\nobreak (1-p_\Psi(t)) \ket{r}\!\bra{r}$.
Therefore, the speedup Hamiltonian  becomes
\begin{align}
\label{eq:app-HspeedupV2}
H_t^{\qs} &\coloneqq \sign(z_t) \lambda_t \frac{i}{2}  \sum_{j \neq k}  \frac{(p_j + p_k)}{p_j - p_k} A_{jk} \ket{j}\!\bra{k} \nonumber \\
&=  \sign(z_t) \lambda_t \frac{i}{2} \left(  \frac{ p_\Psi + 0 }{p_\Psi - 0 }   \bra{\Psi_t}\sigma_z\ket{\Psi_t^\perp} \ket{\Psi_t}\!\bra{\Psi_t^\perp} + \frac{ 0 + p_\Psi  }{ 0 - p_\Psi }   \bra{\Psi_t^\perp}\sigma_z\ket{\Psi_t} \ket{\Psi_t^\perp}\!\bra{\Psi_t} \right)   \nonumber \\ 
&=  \sign(z_t) \lambda_t \frac{i}{2} \left(    \bra{\Psi_t}\sigma_z\ket{\Psi_t^\perp} \ket{\Psi_t}\!\bra{\Psi_t^\perp} - \bra{\Psi_t^\perp}\sigma_z\ket{\Psi_t} \ket{\Psi_t^\perp}\!\bra{\Psi_t}
 \right),
\end{align}
where $A = \sigma_z$, and we have chosen the sign relative to Eq.~\eqref{eq:app-Hspeedup} to ensure that the coherent dynamics drives the system with $\dot z_t/z_t \leq 0$, i.e., helps hide the fact that $z_0 \neq 0$ by driving $\langle \sigma_z \rangle$ towards $0$. 
Indeed, with the chosen sign in Eq.~\eqref{eq:app-HspeedupV2}, we have $L_C^\qs \nobreak = - \sign(z_t) \nobreak \lambda_t \,  \sigma_{(z,C)}$, and the incoherent drive enforces 
\begin{align}
\label{eq-app:changeinz} 
 \dot z_C \nobreak = \nobreak \cov(\sigma_{(z,C)},\slogq^\qs) \nobreak = \nobreak -\sign(z_t) \frac{\info^C}{\lambda_t}.
 \end{align}

Expressing the qubit states as $\ket{\Psi_t} = a_t\ket{0}+b_t\ket{1}$ and $\ket{\Psi_t^\perp} = b_t \ket{0}-a_t \ket{1}$, we have 
\begin{align}
x_t  &\coloneqq \bra{\Psi_t} \sigma_x\ket{\Psi_t} = 2 a_t b_t, \\
\bra{\Psi_t^\perp}\sigma_z\ket{\Psi_t } &= \bra{\Psi_t}\sigma_z\ket{\Psi_t^\perp} = -2 a_t b_t = - x_t.
\end{align}
Then, 
\begin{align}
H_t^{\qs} &=  \sign(z_t) \lambda_t \frac{i}{2} \left( -   x_t    \ket{\Psi_t}\!\bra{\Psi_t^\perp} + x_t   \ket{\Psi_t^\perp}\!\bra{\Psi_t}
 \right)  =  - \sign(z_t) \lambda_t \frac{i}{2} x_t  \big(      \ket{\Psi_t}\!\bra{\Psi_t^\perp} - \ket{\Psi_t^\perp}\!\bra{\Psi_t}
 \big) \nonumber \\
 &=  - \sign(z_t) \lambda_t \frac{i}{2} x_t  \big(     - \left( a_t^2 + b_t^2 \right) \ket{0}\!\bra{1} + \left( a_t^2 + b_t^2 \right)  \ket{1}\!\bra{0}
 \big)  = - \sign(z_t) \lambda_t \frac{x_t }{2} \left(   - i \ket{0}\!\bra{1} +  i\ket{1}\!\bra{0}
 \right) \nonumber \\
 &= \sign(z_t) \lambda_t \frac{x_t }{2} \sigma_y.
\end{align}
The coefficient $\lambda_t$ is to be defined by the resources available to drive our coherent dynamics. If, for instance, we set the spectral norm of the Hamiltonian to satisfy $\left\|H_t^\qs \right\| = \epsilon$, we obtain $\lambda_t = \frac{2\epsilon}{|x_t|}$.
Then, we find that the normalized optimal Hamiltonian becomes
\begin{align}
H_t^{\qs} &=  \epsilon \sign(z_t)  \frac{x_t}{|x_t|} \sigma_y  =  \epsilon \, \frac{\sign(z_t)}{  \sign(x_t)} \, \sigma_y. 
\end{align}
At any time, this Hamiltonian enhances the hiding process as much as allowed by coherent dynamics.

With this, we can compare the rates of erasure of the incoherent and the coherently-enhanced dynamics. At any time $t$, the coherently-enhanced state evolves following
\begin{align}
 \label{eq-app:enhancedchannel}
 \dot \rho_t &= - i [H_t^{\qs},\rho_t] + \gamma \left( \tr{\rho_t \Pi_{2}}   \ket{r}\!\bra{r}  - \frac{1}{2} \{ \Pi_2 , \rho_t  \}  \right).
 \end{align}
 Thus, the rate of change in $z_t$ satisfies
 \begin{align}
     \dot z_{t}  &= -i \tr{ \big[\sigma_z, H_t^\qs 
    \big] \rho_t } - \gamma \, z_t   = -i \epsilon \, \frac{\sign(z_t)}{  \sign(x_t)} \, \tr{ [\sigma_z, \sigma_y 
     ] \rho_t } - \gamma \, z_t   \nonumber \\
     &= - 2   \epsilon \, \frac{\sign(z_t)}{  \sign(x_t)} \,  \langle \sigma_x \rangle   - \gamma \, z_t,
 \end{align}
 which means that
 \begin{align}
     \dot z_t / z_t  = - \gamma - 2 \epsilon \, \left| \frac{x_t}{z_t}  \right|.
 \end{align}
 This process is faster than the one from the purely incoherent erasure process in Eq.~\eqref{eq-app:incoherasure}, $\dot z_t^{\textnormal{incoh}} = - \gamma \, z_t^{\textnormal{incoh}}$. This proves the claim in the two paragraphs following Eq.~\eqref{eq:Qspeed} of Sec.~\ref{sec:geometry} in the main text.

\section{Limits to speed in Hilbert space}
\label{app:qslHilbert}

In this Appendix, we derive speed limits in state space from the coherent-incoherent bounds on observables, placing upper bounds on the rate of change of the quantum fidelity $F(\rho_t,\rho_0)$ between initial and evolved state, referred to after Eq.~\eqref{eq:buresmetric} in the main text and Footnote~\cite{foonote1} in the main text.

\

We assume a pure initial state, $\rho_{0}^2 = \rho_{0} = \chi_{0}$. In this case, the fidelity between the initial and evolved states becomes $F_t \coloneqq F(\rho_t,\rho_0) \coloneqq \left( \tr{\sqrt{ \sqrt{\rho_t} \rho_0 \sqrt{\rho_t} } } \right)^2 = \tr{\rho_0 \rho_t}$.
Taking $A = \rho_0$ in Eqs.~\eqref{eq:bound2} and~\eqref{eq:bound2v2}, $F_t$  satisfies 
\begin{align}
\left|\dt F_t \right| &=\left| \tr{\rho_0 \dt \rho_t} \right| = \left| \cov(\rho_C,\slogq) + \cov(\rho_I,\slogc) \right|  \leq  \Delta \rho_C  \,\sqrt{\info^C} + \Delta \rho_I \, \sqrt{\info^I},
\end{align}
where 
\begin{align}
\rho_C = \sum_{j\neq k} \rho_{0,jk} \ket{j}\!\bra{k}, \qquad \rho_I = \sum_{j} \rho_{0,j} \ket{j}\!\bra{j}
\end{align}
are the coherent and incoherent components of the initial state $\rho_0$ in the eigenbasis of the evolved state $\rho_t$. 

From Appendix~\ref{app:comparison}, we also have 
\begin{align}
\left|\dt F_t \right| &\leq  \Delta \rho_C  \,\sqrt{\info^C} + \Delta \rho_I \, \sqrt{\info^I} \leq  \Delta \rho_0  \,\sqrt{\info},
\end{align}
with a weaker bound in terms of the Fisher information $\info$.

Using 
\begin{align}
(\Delta \rho_C)^2 &= (\Delta \rho_0)^2 - (\Delta \rho_I)^2,
\\
(\Delta \rho_0)^2 &= \tr{\rho_t \rho_0^2} - (\tr{\rho_t \rho_0})^2 = F_t - F_t^2,
\end{align}
we obtain
\begin{align}
\frac{\big|\dt F_t \big|}{\sqrt{F_t - F_t^2}} &\leq  \sqrt{1 - \frac{(\Delta \rho_I)^2}{F_t - F_t^2}}  \,\sqrt{\info^C} + \sqrt{ \frac{(\Delta \rho_I)^2}{F_t - F_t^2} } \, \sqrt{\info^C} \leq \sqrt{\info}.
\end{align}
Upon integration, this gives a bound on the total change in fidelity:
\begin{align}
\arccos(\sqrt{F(\rho_\tau,\rho_0)}) \leq \int_0^\tau \sqrt{1 - \frac{(\Delta \rho_I)^2}{F_t - F_t^2}}  \,\sqrt{\frac{\info^C}{4}} + \sqrt{ \frac{(\Delta \rho_I)^2}{F_t - F_t^2} } \, \sqrt{\frac{\info^C}{4}} dt               \leq \int_0^\tau\sqrt{\frac{\info}{4}} dt.
\end{align}
The rightmost bound is the one derived in~\cite{DavidovichPRL2013}. The tighter intermediate bound is made possible by singling out the coherent and incoherent effects on the change in the fidelity. The two bounds coincide only for purely coherent dynamics of a quantum system.

\section{Comparison to bounds from parameter estimation theory }
\label{app:ParamEst}

In this Appendix, we compare bounds~\eqref{eq:bound2}, and~\eqref{eq:bound2v2} in Secs.~\ref{sec:generalQSL} and~\ref{sec:coherent-incoherent} of the main text to previously known bounds from parameter estimation theory.
We show how to recover Eq.~\eqref{eq:generalbound} in the main text from the Cram\'er-Rao bound.

 \subsection{The quantum Cram\'er-Rao bound}
 \label{app:ParamEstQCR}
  
The quantum Cram\'er-Rao bound generalizes the Cram\'er-Rao bound from classical estimation theory to quantum systems~\cite{HELSTROM1967,helstrom1969quantum,
 holevo2011probabilistic,
 BraunsteinCaves1994}. 
When estimating a parameter $\lambda$ on a system in state $\rho(\lambda)$, the standard deviation of any estimator $\hat \lambda$ of the parameter $\lambda$ satisfies~\cite{BraunsteinCaves1994}
 \begin{align}
 \label{eq-app:QCramerRao}
\frac{ \Delta \hat \lambda  }{\left| \tfrac{d \, }{ d \lambda} \langle \hat \lambda \rangle \right|} \geq \frac{1}{ \sqrt{\info} }.
 \end{align}
The quantum Fisher information is
\begin{align}
\info \coloneqq 2 \sum_{jk}  \frac{\left| \bra{j} \frac{\partial \rho(\lambda)}{\partial \lambda} \ket{k}\right|^2}{p_j + p_k},
\end{align}
with a summation 
 over indexes such that $p_j\nobreak+\nobreak p_k\nobreak\neq\nobreak 0$.
The bound is achievable for the optimal estimator~\cite{BraunsteinCaves1994}. 
Note that the bound assumes that the estimator $\hat \lambda$ is independent of the parameter $\lambda$ to be estimated~\cite{BraunsteinCaves1994,paris2009quantum,sidhu_geometric_2020} (we discuss this further below).

The quantum parameter estimation problem involves two optimizations:
(i) optimizing over all possible observables that can be measured---more generally, optimizing over all possible positive operator valued measures (POVMs) that can be performed--and
(ii) optimizing over all possible estimators $\hat \lambda$ that can be constructed from the measurement outcomes. 
Optimization (ii) is accounted for by the classical Cram\'er-Rao bound, but (i) entails a purely quantum aspect to the problem.

\subsection{Restricting the estimator to functions of $\langle A \rangle$}
 \label{app:ParamEstGeneralBound}

In this paper, we focus on speed limits, i.e., on the rate of change of the expectation value of an observable $A$. The Cram\'er-Rao bound can also be cast as a bound on this rate of change.

Focusing on the case of 
time as the parameter $\lambda = t$  to be estimated, Eq.~\eqref{eq-app:QCramerRao} imposes a bound on the rate of change of the mean of any estimator $\hat t$ of $t$ as. If we restrict 
to an observable $A = \hat{t}$, the Cram\'er-Rao bound implies
\begin{align}
\label{eq-app:CRobservable}
\left| \frac{d}{dt} \left\langle A \right\rangle  \right| \leq \sqrt{\info} \Delta A.
\end{align}


For the case of operators without explicit time dependence, bound~\eqref{eq:generalbound} in the main text coincides with  bound~\eqref{eq-app:CRobservable} implied by the quantum Cram\'er-Rao theorem.    
The Cram\'er-Rao bound assumes no time parameter dependence in the estimators~\cite{BraunsteinCaves1994} though, so it does not directly recover~\eqref{eq:generalbound} for time-dependent operators.
 However, the following procedure allows time-dependent operators.
We wish to find a speed limit at time $t= t_0$ for the expectation value $\tr{\rho_t A(t)}$ of an operator $A(t)$ (to avoid confusion, we write the explicit time dependence for the  proof that follows). 
Bound~\eqref{eq-app:CRobservable} applies to any operator, and in particular to $A(t_0)$, and implies
\begin{align}
\left| \frac{d}{dt} \tr{\rho_t A(t_0)} \right| \leq \sqrt{\info(t)} \Delta A(t_0).
\end{align}
Evaluating this bound at $t = t_0$ recovers  bound~\eqref{eq:generalbound} in the main text.

\subsection{Identifying coherent and incoherent contributions to the dynamics}
\label{app:ParamEstCohIncohBound}

The restriction to a specific observable and the specification of the evolution of the state that singles out contributions from unitary and incoherent dynamics enable the main bounds in this article: upper bounds~\eqref{eq:bound2} and~\eqref{eq:bound2v2}. 
As we prove in Appendix~\ref{app:comparison}, these novel bounds are tighter than bound~\eqref{eq-app:CRobservable} derivable from the quantum Cram\'er-Rao bound. 

Moreover, we stress that the lower speed bounds~\eqref{eq:boundlower} are not accounted for by Cram\'er-Rao bounds, but are instead made possible by the separation of the dynamics into two terms, which in turn allows for applying the reverse triangle inequality.

\subsection{An alternative bound with a basis-dependent classical Fisher information}
 \label{app:ParamEstCCR}

A possible parameter estimation problem is to forego optimization (i) in the quantum Cram\'er-Rao bound above, and instead restrict to a specific measurement basis.
 Then, for a fixed set of POVMs $\{ \Pi_\alpha \}$ with outcome probabilities $q_\alpha^{\Pi} = \tr{\rho \Pi_\alpha}$, this recovers a classical parameter estimation problem. 
 One can directly obtain a classical Fisher information from the probabilities $\{ q^{\Pi}_\alpha\}$: 
\begin{align}
\label{eq-app:classicalFisher}
F_{\Pi} \coloneqq \sum_\alpha q^{\Pi}_\alpha \left( \frac{d}{dt} \ln q^{\Pi}_\alpha \right)^2.
\end{align} 
The classical Cram\'er-Rao bound thus says that, for any estimator $\hat t$ constructed from 
outcomes of the POVM $\{ \Pi_\alpha \}$, it holds that~\cite{darmois1945limites,
cramir1946mathematical,
rao1992information}
\begin{align}
\left| \frac{d}{dt} \langle \hat t \rangle  \right| \leq \sqrt{ F_{\Pi} \Delta \hat t }. 
\end{align}
 If measurements are performed solely in the restricted measurement basis, this bound is tighter than the quantum Cram\'er-Rao bound~\eqref{eq-app:QCramerRao}, since the latter is valid for any measurement basis. Bound~\eqref{eq-app:classicalFisher} involves optimization (ii) over all possible estimators, given the restricted measurement basis $\{\Pi_\alpha \}$.

 Further restricting to $\langle A \rangle$  as an estimator gives
 
 \begin{align}
 \label{eq-app:CCramerRao}
\left| \frac{d}{dt} \langle A \rangle  \right| \leq \sqrt{ F_{\Pi}  } \Delta A, 
\end{align}
 which is in fact \emph{tighter} than~\eqref{eq-app:CRobservable}, since $F_\Pi \leq \info$.
However,  
note that the classical Fisher information~\eqref{eq-app:classicalFisher}  
depends on the measurement basis, too. In general, then, calculating $F_\Pi$ is more intricate, and dependent on the system dynamics, the state, \emph{and} a reference measurement basis.
In contrast, $\info$ (as well as $\info^C$ and $\info^I$) are functions only of the state of the system and the dynamics that govern it.

When the measurement basis is chosen as the eigenbasis of the state of the system, $\{ \Pi_\alpha \} \equiv \{ \ket{j}\!\bra{j} \}$, one obtains $\{ q_\alpha^\Pi \} \equiv \{ p_j \}$, so that the 
classical basis-dependent Fisher information $F_\Pi$ coincides with $\info^I$:
\begin{align}
F_{\Pi} = \sum_j p_j \left( \frac{d}{dt} \ln p_j \right)^2 = \info^I.
\end{align}
Then, if the estimator for $t$ is taken to be $\hat t = A_I = \sum_j A_{jj} \ket{j}\!\bra{j}$, the classical Cram\'er-Rao bound~\eqref{eq-app:CCramerRao} recovers the bound on the incoherent term in~Eq.~\eqref{eq:bound2} of the main text:
\begin{align}
\left| \frac{d}{dt} \langle A_I \rangle  \right| \leq \sqrt{ \info^I } \Delta A_I,
\end{align} 
 for time-independent operators (which can be extended to time-dependent operators as was shown in Appendix~\ref{app:ParamEstGeneralBound}). 
  
However, a similar trick to restrict the basis does not work on the bound for the coherent term $A_C$, since the eigenbasis of $A_C$ does not commute with that of $\rho$. A classical parameter estimation bound on $A_C$ would then yield a basis-dependent classical Fisher information.

This discussion results in a hierarchy of bounds
 based on the level of optimization involved in them and their tightness. The main ones in this article--Eqs.~\eqref{eq:bound2},~\eqref{eq:bound2v2}, and~\eqref{eq:boundlower}--are tailored to the problem of the speed of evolution of an observable, and as a result are the tightest among all the bounds considered. In contrast, Cram\'er-Rao bounds focus on a different (more general) question, and are looser as a result.


\end{document}